\documentclass{article}

\usepackage{xcolor}
\usepackage{amsmath}
\usepackage{graphicx}
\usepackage{subfig}
\usepackage{hyperref}
\usepackage{cleveref}
\usepackage{natbib}
\usepackage{algorithm}
\usepackage{algpseudocode}
\usepackage[margin=1in]{geometry}
\graphicspath{{./figures/}}

\begin{document}

\title{A 3D VTI factored eikonal solver using six-tetrahedron pyramidal stencil\thanks{Published as: Yuhang Wang, Yongming Lu, Jianming Zhang, and Pengliang Yang, ``A 3D VTI Factored Eikonal Solver Using a Six-Tetrahedron Pyramidal Stencil,'' \textit{Geophysical Prospecting}, 2026, Article ID GPR70250. DOI: \href{https://doi.org/10.1111/1365-2478.70250}{10.1111/1365-2478.70250}. Accepted 13 August 2026.}}

\author{
  Yuhang Wang$^{1}$ \and
  Yongming Lu$^{2}$ \and
  Jianming Zhang$^{3}$ \and
  Pengliang Yang$^{4}$\thanks{Corresponding author. E-mail: ypl.2100@gmail.com}
}
\date{\today}

\maketitle

\begin{center}
  \footnotesize
  $^{1}$School of Mathematics, Harbin Institute of Technology, Harbin 150001, China\\
  $^{2}$Shenzhen MSU-BIT University, Shenzhen 518172, China\\
  $^{3}$Ocean University of China, Qingdao 266100, China\\
  $^{4}$Laoshan Laboratory, Qingdao, 266237, China
\end{center}

\begin{abstract}
  Accurate traveltime computation for the eikonal equation is essential in seismic applications such as tomography and migration. The fast sweeping method (FSM) is widely used because of its unconditional stability and computational efficiency. We develop a 3D fast sweeping solver for vertical transverse isotropic (VTI) media that combines multiplicative factorization with a six-tetrahedron pyramidal stencil. The factorization removes the point-source singularity, while the stencil improves local accuracy. In the unfactored formulation, the six-tetrahedron finite-difference scheme requires solving only quadratic equations. After factorization, however, the local update for the perturbation factor becomes quartic, and the update systems on oblique stencil faces are substantially more complicated than those on non-oblique faces. To resolve these difficulties, we solve the update equation using Ferrari's method with a robust root-selection strategy  and derive complete update formulas for all oblique-face configurations. For horizontally constrained faces, the characteristic constraint reduces to a linear relation by exploiting the structure of the VTI Hamiltonian, so the local system still reduces to a quartic equation. For mixed horizontal-vertical constrained faces, we design a bisection-based iterative solver. Numerical examples show that the proposed method effectively suppresses source-related errors and improves traveltime accuracy.
\end{abstract}


\section{Introduction}

The eikonal equation plays a central role in wave propagation, geometrical optics, and exploration seismology because it describes the spatial variation of wave traveltime in a medium with prescribed velocity or slowness. In seismology, first-arrival traveltimes  are essential for seismic ray tracing \citep{Cerveny}, velocity model building and tomography \citep{zhoubing2008,Taillandier2009,Huangjw2012,umair2016tomo}, migration imaging \citep{Gray1994,Simon2019}, and microseismic event location \citep{Grechka2015,zhang2023eikonal,Gao2026LATTE}. Accurate traveltime computation is especially important in anisotropic media, where wave propagation depends strongly on direction. As a high-frequency asymptotic approximation of the full wave equation, the eikonal equation neglects amplitude information and focuses on wavefront kinematics. Its strong nonlinearity, together with anisotropy, makes the design of accurate, stable, and efficient numerical solvers particularly challenging.

Finite-difference (FD) discretization is one of the most common approaches for eikonal-based traveltime computation. Among the available methods, the fast marching method (FMM) \citep{Sethian1996A,sethian19993d} and the fast sweeping method (FSM) \citep{Zhao_2005_FSM,Zhang2006,qian2007f} are the most widely used. Compared with conventional ray tracing, grid-based FD methods are generally more robust in complex heterogeneous media because they avoid ray-path singularities and provide full coverage of the computational domain on a regular mesh. FSM is particularly attractive because it is unconditionally stable, straightforward to implement, and computationally efficient. It well adapts to anisotropic problems, including the vertical transverse isotropic (VTI) media commonly encountered in seismic exploration.

FSM uses a Gauss-Seidel-type iteration with alternating sweeps along different characteristic directions. This multidirectional sweeping strategy naturally incorporates wave propagation from all directions and therefore adapts well to anisotropic eikonal equations without requiring substantial additional algorithmic machinery \citep{zhanglj2018,Lu2025}. Moreover, the iterative structure of FSM is well matched to the nonlinearity of anisotropic eikonal systems, enabling accurate traveltime computation in media with strong parameter variations while maintaining a relatively low memory cost.

Considerable effort has been devoted to extending FSM to anisotropic media. Early studies \citep{lan2014,Waheed2015A,Han2017,Le2018,Le2019,Huang2020,2020Hybrid,zhou2021,zhangqy2021} established the foundation for efficient traveltime computation in direction-dependent velocity models. Later work \citep{Rawlinson2004MultipleRA,Fomel_2009_FSM,Luo2015High,Waheed2015F} identified two major obstacles that limit both accuracy and efficiency in practice. The first is the point-source singularity, which often causes large numerical errors near the source. The second is the algebraic complexity of anisotropic local updates, which frequently leads to fourth-order polynomial equations. These high-order equations increase the cost of local solves and complicate the selection of physically admissible roots.

Factored eikonal formulations are an effective way to suppress source singularities \citep{Fomel_2009_FSM}. At the same time, high-accuracy stencils have been developed to improve the local approximation of anisotropic wave propagation. However, combining these two ideas is not straightforward. In VTI media, introducing factorization into a high-accuracy stencil turns the local update for the perturbation factor into a quartic equation with complicated coefficients, and the corresponding update systems on oblique stencil faces become substantially more difficult to solve. As a result, the combination of source-singularity removal and improved local accuracy introduces a new computational bottleneck.

Several studies have addressed different aspects of this problem. \citet{Luo_2012_FSM} proposed a factorization strategy that removes the source singularity and improves accuracy near the source. \citet{Waheed2015Efficient} developed a fixed-point iteration scheme that improves the efficiency of anisotropic FSM updates. 
More recently, \citet{lu2021fast,Lu2025} introduced a six-tetrahedron finite-difference stencil tailored to the directional propagation characteristics of VTI and TTI media. In the unfactored setting, this stencil reduces the local update to a quadratic equation with explicit roots, thereby avoiding the ambiguity and cost associated with quartic solvers. The
stencil is also second-order accurate for smooth models, which makes it a natural foundation for a factored extension. \citet{Huang_2026_3D_VTI_FSM} compute the group velocity with application to traveltime computation in 3D VTI media using FSM, although the source singularity persists in their approach.

The motivation of this work is to combine the advantages of multiplicative factorization and the six-tetrahedron pyramidal stencil within a single 3D VTI solver. To achieve this goal, we develop an FSM algorithm for the 3D VTI eikonal equation that integrates these two ingredients while explicitly addressing the additional algebraic complexity introduced by factorization. We solve the resulting quartic equations using Ferrari's method and systematically enforce causality constraints. For oblique stencil faces, we derive update strategies for both horizontally constrained and mixed horizontal-vertical constrained configurations. In this way, the proposed method removes the source singularity while preserving the improved local accuracy of the six-tetrahedron stencil.

\section{Method}

\subsection{3D VTI Eikonal Equation and FSM Discretization}

Let $\mathbf{p}=(p_x,p_y,p_z)=\left(\frac{\partial T}{\partial x}, \frac{\partial T}{\partial y}, \frac{\partial T}{\partial z}\right)$ denote the slowness vector associated with the traveltime field $T$. The eikonal equation for VTI media is \citep{Alkhalifah_1998_AAP}
\begin{equation}\label{eq:eikonal}
  V_{\text{nmo}}^{2}(1+2\eta)(p_{x}^{2}+p_{y}^{2})+V_{0}^{2}p_{z}^{2}\bigl(1-2\eta V_{\text{nmo}}^{2}(p_{x}^{2}+p_{y}^{2})\bigr)=1,
\end{equation}
where $V_{\text{nmo}}$ and $V_0$ denote the NMO velocity and the vertical velocity, respectively, and $\eta=(\epsilon-\delta)/(1+2\delta)$ is the anisotropy parameter defined in terms of Thomsen's parameters $\epsilon$ and $\delta$.

Denote the left-hand side of \eqref{eq:eikonal} by the Hamiltonian $H(\mathbf{p})$. The characteristic directions are determined by $\nabla_{\mathbf{p}}H=\left(\frac{\partial H}{\partial p_x}, \frac{\partial H}{\partial p_y}, \frac{\partial H}{\partial p_z}\right)$:
\begin{subequations}\label{eq:partial_derivatives}
  \begin{align}
    \frac{\partial H}{\partial p_x} &= 2V_{\text{nmo}}^2 p_x \left(  1 + 2\eta - 2V_0^2  \eta p_z^2 \right), \label{eq:dH_dpx} \\
    \frac{\partial H}{\partial p_y} &= 2V_{\text{nmo}}^2 p_y \left(  1 + 2\eta - 2V_0^2  \eta p_z^2 \right), \label{eq:dH_dpy} \\
    \frac{\partial H}{\partial p_z} &= 2V_0^2 p_z \left( 1 - 2V_{\text{nmo}}^2 \eta (p_x^2 + p_y^2) \right). \label{eq:dH_dpz}
  \end{align}
\end{subequations}

In the finite-difference discretization, the neighborhood of each grid point is divided into eight octants, and each octant is further partitioned into six tetrahedra. \Cref{fig:images-1} shows the decomposition in one octant; the other octants are treated similarly. This decomposition yields six
possible one octant; the other octants are treated in the same way update stencils per octant, each using three known neighbor values to update the unknown at the octant vertex.

\subsubsection{Solution Within Tetrahedral Elements}

\begin{figure}[htbp]
  \centering
  \subfloat[]{\includegraphics[width=0.33\linewidth]{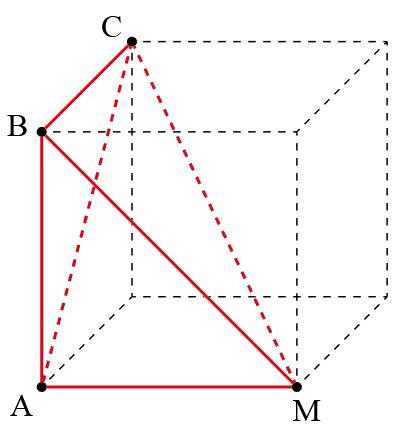}\label{fig:img1}}
  \subfloat[]{\includegraphics[width=0.33\linewidth]{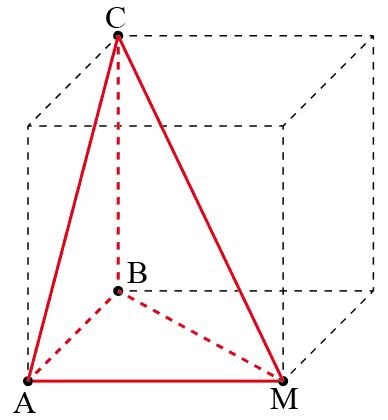}\label{fig:img2}}
  \subfloat[]{\includegraphics[width=0.33\linewidth]{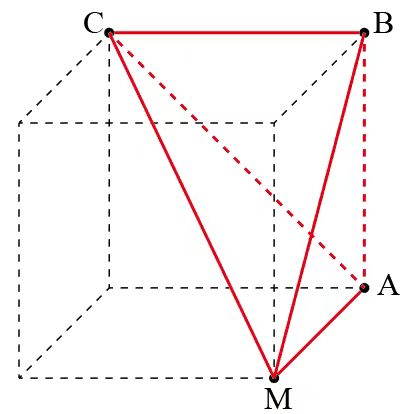}\label{fig:img3}}\\
  \subfloat[]{\includegraphics[width=0.33\linewidth]{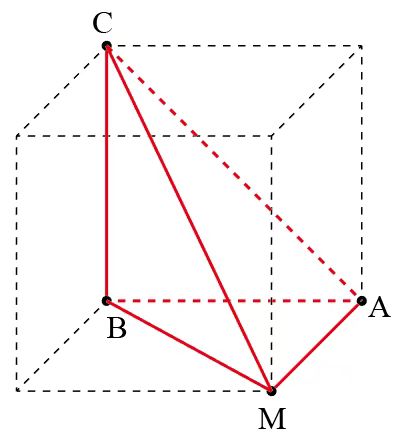}\label{fig:img4}}
  \subfloat[]{\includegraphics[width=0.33\linewidth]{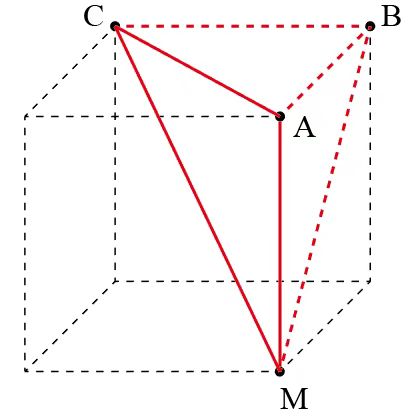}\label{fig:img5}}
  \subfloat[]{\includegraphics[width=0.33\linewidth]{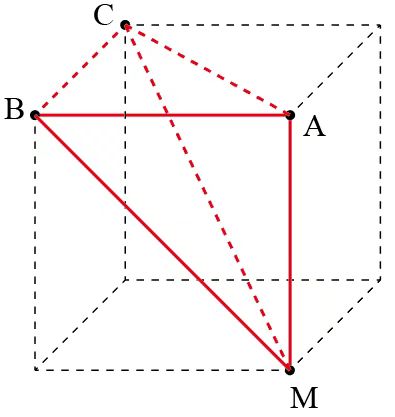}\label{fig:img6}}\\
  \caption{Tetrahedral mesh decomposition within a single octant (The traveltime at point $M$ is the unknown to be determined).}  \label{fig:images-1}
\end{figure}

In the tetrahedron shown in \Cref{fig:img1}, we define the spatial increments
\[
dx = x_M - x_A, \quad dy = y_C - y_B, \quad dz = z_B - z_A,
\]
so that the traveltime differences along $MA$, $MB$, and $MC$ can be approximated by
\begin{subequations}
  \begin{align}
    T_A - T_M &= -dx \, p_x, \\
    T_B - T_M &= -dx \, p_x + dz \, p_z, \\
    T_C - T_M &= -dx \, p_x + dy \, p_y + dz \, p_z.
  \end{align}
\end{subequations}
Solving these relations gives
\begin{subequations}\label{eq:difference}
  \begin{align}
    p_x &= \frac{T_M - T_A}{dx}, \label{eq:difference.a}\\
    p_y &= \frac{T_C - T_B}{dy}, \label{eq:difference.b}\\
    p_z &= \frac{T_B - T_A}{dz}. \label{eq:difference.c}
  \end{align}
\end{subequations}
Among these quantities, only $p_x$ remains unknown. Substituting \eqref{eq:difference} into \eqref{eq:eikonal} yields a quadratic equation for $p_x$. Once this equation is solved, the corresponding traveltime update follows from back-substitution into \eqref{eq:difference}.

This procedure may yield up to two real roots. Each root is checked by tracing the characteristic backward from point $M$ and verifying that it intersects the interior of triangle $ABC$. If two roots satisfy this causality condition, we take the smaller one; if only one root is valid, we use it directly. If no valid root is found, we compute the updates on the three tetrahedral faces passing through $M$ and take the smallest valid value.

The configuration in \Cref{fig:img2} and  \Cref{fig:img1} follow the same pattern. In \Cref{fig:img2}, $p_y$ and $p_z$ are obtained by finite differences, and $p_x$ is recovered from the eikonal equation.  In \Cref{fig:img3} and \Cref{fig:img4}, $p_x$ and $p_z$ are computed by finite differences and $p_y$ is solved from the eikonal equation. In \Cref{fig:img5} and \Cref{fig:img6}, $p_x$ and $p_y$ are computed by finite differences and $p_z$ is determined from the eikonal equation.

\subsubsection{Solution on Tetrahedral Faces}
\begin{figure}[htbp]
  \centering
  \subfloat[]{\includegraphics[width=0.33\linewidth]{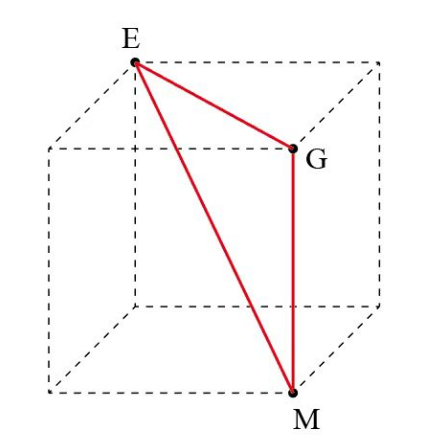}\label{fig:img2.1}}
  \subfloat[]{\includegraphics[width=0.33\linewidth]{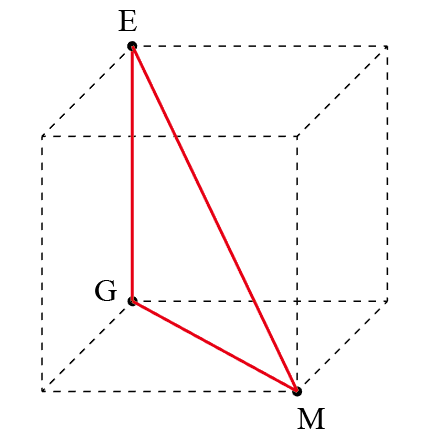}\label{fig:img2.2}}
  \subfloat[]{\includegraphics[width=0.33\linewidth]{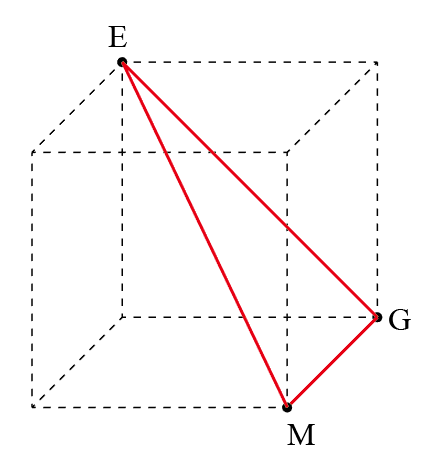}\label{fig:img2.3}}\\
  \subfloat[]{\includegraphics[width=0.33\linewidth]{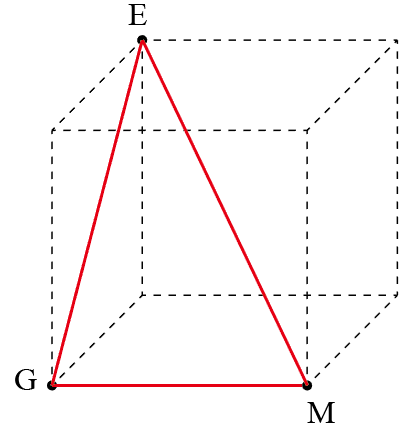}\label{fig:img2.4}}
  \subfloat[]{\includegraphics[width=0.33\linewidth]{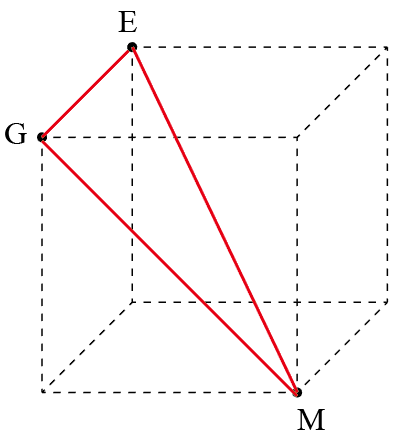}\label{fig:img2.5}}
  \subfloat[]{\includegraphics[width=0.33\linewidth]{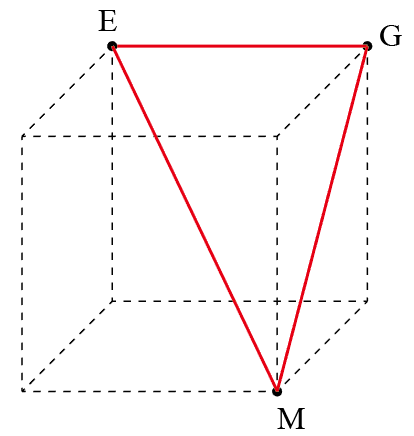}\label{fig:img2.6}}\\
  \caption{Characteristic configurations on inclined planes.}\label{fig:images-2}
\end{figure}

We next consider the case in which the characteristic lies on a tetrahedral face. Axis-aligned faces are straightforward; inclined faces require an additional characteristic constraint. \Cref{fig:images-2} shows the representative inclined-face configurations.

Let $\mathbf{n}$ be a normal vector to the face plane. Requiring the characteristic direction to remain on that plane is equivalent to enforcing orthogonality between $\nabla_{\mathbf{p}}H$ and $\mathbf{n}$, namely
\[
\mathbf{n} \cdot \nabla_{\mathbf{p}} H = 0,
\]
which we use as the characteristic constraint equation.

\paragraph{Case 1: Horizontal Characteristic Constraint}
For the configuration shown in \Cref{fig:img2.1}, we define
\[
dx = x_E - x_G, \quad dy = y_E - y_G, \quad dz = z_M - z_G.
\]
Let $\mathbf{n} = (-dy, dx, 0)$ be a normal vector to triangle $EGM$. Enforcing $\mathbf{n} \cdot \nabla_\mathbf{p} H = 0$ gives the characteristic constraint for this face. The traveltimes then satisfy
\begin{subequations}
  \begin{align}
    T_E - T_G &= p_x \, dx + p_y \, dy, \\
    T_M - T_G &= p_z \, dz, \\
    -dy \frac{\partial H}{\partial p_x} + dx \frac{\partial H}{\partial p_y} &= 0.
  \end{align}
\end{subequations}
Hence,
\begin{subequations}
  \begin{align}
    p_x &= \frac{(T_E - T_G)dx}{dx^2 + dy^2}, \\
    p_y &= \frac{(T_E - T_G)dy}{dx^2 + dy^2}.
  \end{align}
\end{subequations}
Substituting these expressions into \eqref{eq:eikonal} yields the local update, which is then checked against the same causality condition as in the interior tetrahedral case. If the test fails, we compute the one-segment updates along $\overrightarrow{EM}$ and $\overrightarrow{GM}$ and retain the smaller valid value. The update can be written as
\[
T_M^{\text{new}} =
\begin{cases}
  \min\left(T_M^{\text{old}}, T_1^{\text{valid}}, T_2^{\text{valid}}\right), & \text{if two valid roots } T_1^{\text{valid}}, T_2^{\text{valid}} \text{ exist}, \\[6pt]
  \min\left(T_M^{\text{old}}, T^{\text{valid}}\right), & \text{if only one valid root } T^{\text{valid}} \text{ exists}, \\[6pt]
  \min\left(T_M^{\text{old}}, T_E + \dfrac{|\overrightarrow{EM}|}{v_{\overrightarrow{EM}}}, T_M + \dfrac{|\overrightarrow{GM}|}{v_{\overrightarrow{GM}}}\right), & \text{if no valid root exists},
\end{cases}
\]
where $v_{\overrightarrow{EM}}$ and $v_{\overrightarrow{GM}}$ are the group velocities along the characteristic directions 
$\overrightarrow{EM}$ and $\overrightarrow{GM}$, respectively. 

The configuration in \Cref{fig:img2.2} is treated analogously. There, $p_z$ is obtained by finite differences, while $p_x$ and $p_y$ are recovered from the eikonal equation under the same proportionality relation as in \Cref{fig:img2.1}.

\paragraph{Case 2: Vertical--Horizontal Mixed Constraint}
For \Cref{fig:img2.3}, the spatial increments are
\[
dx = x_E - x_G, \quad dy = y_M - y_G, \quad dz = z_E - z_G,
\]
and let $\mathbf{n} = (-dz, 0, dx)$ be a normal vector to triangle $EGM$. The traveltimes then satisfy
\begin{subequations}\label{eq:system1}
  \begin{align}
    T_E - T_G &= p_x \, dx + p_z \, dz, \label{eq:system1.a}\\
    T_M - T_G &= p_y \, dy, \label{eq:system1.b}\\
    -dx \frac{\partial H}{\partial p_z} + dz \frac{\partial H}{\partial p_x} &= 0. \label{eq:system1.c}
  \end{align}
\end{subequations}

We solve this system by bisection. Define the residual
\[
f(T_M)=T_E-T_G-p_x\,dx-p_z\,dz,
\]
where $p_x$ and $p_z$ are evaluated from \eqref{eq:system1}. Under the physical constraint $|p_z| \in [0, 1/V_0]$, each trial value of $p_z$ determines $p_x^2+p_y^2$ through \eqref{eq:eikonal}; \eqref{eq:system1.c} together with \eqref{eq:dH_dpx} and \eqref{eq:dH_dpz} then gives $p_x$, and \eqref{eq:system1.b} gives $p_y$ and $T_M$. The bisection iteration enforces $f=0$. The resulting traveltime is accepted only if it satisfies the causality condition; otherwise we compare the one-segment updates along $\overrightarrow{EM}$ and $\overrightarrow{GM}$ and keep the smaller valid value. The update can be written as
\[
T_M^{\text{new}} =
\begin{cases}
  \min\left(T_M^{\text{old}}, T^{\text{valid}}\right), & \text{if only one valid root } T^{\text{valid}} \text{ exists}, \\[6pt]
  \min\left(T_M^{\text{old}}, T_E + \dfrac{|\overrightarrow{EM}|}{v_{\overrightarrow{EM}}}, T_M + \dfrac{|\overrightarrow{GM}|}{v_{\overrightarrow{GM}}}\right), & \text{if no valid root exists},
\end{cases}
\]
where $v_{\overrightarrow{EM}}$ and $v_{\overrightarrow{GM}}$ are the group velocities along the characteristic directions $\overrightarrow{EM}$ and $\overrightarrow{GM}$, respectively.

The configuration in \Cref{fig:img2.4} is handled in the same way. One only needs to redefine the geometric increments and interchange $(p_x,dx)$ with $(p_y,dy)$ in \eqref{eq:system1}.

\paragraph{Case 3: Alternative Mixed Constraint}

For \Cref{fig:img2.5}, the spatial parameters are
\[
dx = x_M - x_G, \quad dy = y_E - y_G, \quad dz = z_M - z_G,
\]
and the traveltimes then satisfy
\begin{subequations}\label{eq:system2}
  \begin{align}
    T_E - T_G &= p_y \, dy, \label{eq:system2.a}\\
    T_M - T_G &= p_x \, dx + p_z \, dz, \label{eq:system2.b}\\
    -dx \frac{\partial H}{\partial p_z} + dz \frac{\partial H}{\partial p_x} &= 0. \label{eq:system2.c}
  \end{align}
\end{subequations}

Since $p_y$ is known, the admissible range for $p_x^2$ is
\[
p_x^2 \in \left(0, \frac{1}{V_{\text{nmo}}^2(1+2\eta)} - p_y^2\right).
\]
We again use bisection. Define the residual
\[
f(T_M) = -\frac{\partial H}{\partial p_x}\,dz + \frac{\partial H}{\partial p_z}\,dx,
\]
with $p_x$ and $p_z$ constrained by \eqref{eq:system2}. For each trial value of $p_x$, we recover $p_y$ and $p_z$ from \eqref{eq:eikonal} and \eqref{eq:system2.a}, evaluate $\frac{\partial H}{\partial p_x}$ and $\frac{\partial H}{\partial p_z}$ from \eqref{eq:dH_dpx} and \eqref{eq:dH_dpz}, and then compute $T_M$ from \eqref{eq:system2.b}. The accepted root must satisfy the causality condition; otherwise we compare the one-segment updates along $\overrightarrow{EM}$ and $\overrightarrow{GM}$ and keep the smaller valid value. The update can be written as
\[
T_M^{\text{new}} =
\begin{cases}
  \min\left(T_M^{\text{old}}, T^{\text{valid}}\right), & \text{if only one valid root } T^{\text{valid}} \text{ exists}, \\[6pt]
  \min\left(T_M^{\text{old}}, T_E + \dfrac{|\overrightarrow{EM}|}{v_{\overrightarrow{EM}}}, T_M + \dfrac{|\overrightarrow{GM}|}{v_{\overrightarrow{GM}}}\right), & \text{if no valid root exists},
\end{cases}
\]
where $v_{\overrightarrow{EM}}$ and $v_{\overrightarrow{GM}}$ are the group velocities along the characteristic directions $\overrightarrow{EM}$ and $\overrightarrow{GM}$, respectively.

The configuration in \Cref{fig:img2.6} is treated analogously after interchanging $(p_x,dx)$ with $(p_y,dy)$ and redefining the geometric increments.

\subsection{Source Singularity and Multiplicative Factorization}

The conventional FSM formulation described above does not explicitly remove the point-source singularity, which can degrade solution accuracy near the source. To address this issue, we introduce a multiplicative factorization of the traveltime field:
\begin{equation}
  T = T_0 \, \tau,
  \label{eq:mul_fac}
\end{equation}
where $T_0$ is a known reference field and $\tau$ is the unknown perturbation factor. The corresponding gradient is
\begin{equation}
  \nabla T = \tau \, \nabla T_0 + T_0 \, \nabla \tau.
  \label{eq:grad_fac}
\end{equation}

\subsubsection{Solution Within Tetrahedral Elements}\label{sec:3dvti}

For the configuration in \Cref{fig:img1}, the gradient components of $\tau$ at point $M$ are approximated by
\begin{equation}
  \frac{\partial \tau}{\partial x} = \frac{\tau_M - \tau_A}{dx}, \quad 
  \frac{\partial \tau}{\partial y} = \frac{\tau_C - \tau_B}{dy}, \quad 
  \frac{\partial \tau}{\partial z} = \frac{\tau_B - \tau_A}{dz}.
  \label{eq:diff}
\end{equation}
Here, $\tau_A$, $\tau_B$, and $\tau_C$ are known values at neighboring grid points, whereas $\tau_M$ is the unknown value at point $M$. Consequently, $\frac{\partial \tau}{\partial y}$ and $\frac{\partial \tau}{\partial z}$ can be evaluated directly, while $\frac{\partial \tau}{\partial x}$ remains unknown.

For brevity, we write $\tau=\tau_M$ in the following derivation. Substituting \eqref{eq:diff} into \eqref{eq:grad_fac} gives the components of $\nabla T$ at point $M$:
\begin{subequations}
  \begin{align}
    \frac{\partial T}{\partial x} &= \tau \left(\frac{\partial T_0}{\partial x} + \frac{T_0}{dx}\right) - \frac{T_0 \tau_A}{dx}, \label{eq:grad_Tx} \\
    \frac{\partial T}{\partial y} &= \tau \frac{\partial T_0}{\partial y} + T_0 \frac{\partial \tau}{\partial y}, \label{eq:grad_Ty} \\
    \frac{\partial T}{\partial z} &= \tau \frac{\partial T_0}{\partial z} + T_0 \frac{\partial \tau}{\partial z}. \label{eq:grad_Tz}
  \end{align}
\end{subequations}

Equations \eqref{eq:grad_Tx}--\eqref{eq:grad_Tz} show that each partial derivative of $T$ depends linearly on $\tau$. Substituting these expressions into \eqref{eq:eikonal} therefore yields a quartic equation for $\tau$. For notational convenience, we write
\begin{equation}
  \nabla T = (a_1 \tau + b_1,\ a_2 \tau + b_2,\ a_3 \tau + b_3).
  \label{eq:convenience}
\end{equation}
The eikonal equation then becomes
\begin{equation}
  a \tau^4 + b \tau^3 + c \tau^2 + d \tau + e = 0,
  \label{eq:quartic}
\end{equation}
where the coefficients are
\begin{equation}
  \left\{
  \begin{aligned}
    a &= 2 \eta V_{0}^{2} V_{\text{nmo}}^{2} a_{3}^{2} (a_{1}^{2}+a_{2}^{2}), \\
    b &= 4 \eta V_{0}^{2} V_{\text{nmo}}^{2} a_{3} ((a_{1}^{2}+a_{2}^{2}) b_{3} + (a_1 b_1 + a_2 b_2) a_3 ), \\
    c &= 2 \eta V_{0}^{2} V_{\text{nmo}}^{2} ( (a_{1}^{2}+a_{2}^{2}) b_{3}^{2} + (b_{1}^{2}+b_{2}^{2}) a_{3}^{2} + 4(a_1 b_1 + a_2 b_2) a_3 b_3 )
    - V_{\text{nmo}}^{2}(1+2 \eta) (a_{1}^{2}+a_{2}^{2}) - V_{0}^{2} a_{3}^{2}, \\
    d &= 4 \eta V_{0}^{2} V_{\text{nmo}}^{2} b_{3} ((a_1 b_1 + a_2 b_2) b_3 + (b_{1}^{2}+b_{2}^{2})a_{3} ) 
    - 2 V_{\text{nmo}}^{2}(1+2 \eta) (a_1 b_1 + a_2 b_2) - 2 V_{0}^{2} a_{3} b_{3}, \\
    e &= 1 + 2 \eta V_{0}^{2} V_{\text{nmo}}^{2} ((b_{1}^{2}+b_{2}^{2})b_{3}^{2} ) - V_{\text{nmo}}^{2}(1+2 \eta)(b_1^2 + b_2^2) - V_{0}^{2} b_{3}^{2}. 
  \end{aligned}
  \right.
\end{equation}

Equation \eqref{eq:quartic} is solved by Ferrari's method and may yield up to four real roots. Among these roots, we select the smallest one that satisfies the causality condition. Specifically, a ray traced backward from point $M$ along the associated characteristic direction must intersect the interior of triangle $ABC$. If no valid root is found for the tetrahedral element, we compute the solutions separately on the three tetrahedral faces passing through point $M$ (triangles $ABM$, $ACM$, and $BCM$) and select the smallest valid value.

The updates for \Cref{fig:img2}--\Cref{fig:img6} are analogous to that for \Cref{fig:img1}. The only difference is that the finite-difference formulas for $\tau$ change with the local geometry, which leads to different expressions for $p_x$, $p_y$, and $p_z$, or equivalently for the coefficients $a_1,b_1,a_2,b_2,a_3,b_3$ in \eqref{eq:convenience}. The overall update procedure remains unchanged.

\subsubsection{Solution on Tetrahedral Faces}

When the characteristic line lies on an inclined face, the update procedure differs from that in Section~\ref{sec:3dvti}. We consider these inclined-face cases below.

\paragraph{Case 1: Horizontal Characteristic Constraint}
For the configuration shown in \Cref{fig:img2.1}, we define
\[
dx = x_E - x_G, \quad dy = y_E - y_G, \quad dz = z_M - z_G,
\]
so that
\begin{subequations}
  \begin{align}
    \tau_E - \tau_G &= dx \frac{\partial \tau}{\partial x} + dy \frac{\partial \tau}{\partial y}, \\
    \tau_M - \tau_G &= dz \frac{\partial \tau}{\partial z}.
  \end{align}
\end{subequations}
Substituting these relations into \eqref{eq:grad_fac} gives
\begin{equation}
  \left(dx \frac{\partial T_0}{\partial x} + dy \frac{\partial T_0}{\partial y} \right) \tau + T_0 (\tau_E - \tau_G) = dx p_x + dy p_y,
  \label{eq:diff2.2.1}
\end{equation}
\begin{equation}
  dz \frac{\partial T_0}{\partial z} \tau + T_0 (\tau - \tau_G) = dz p_z.
  \label{eq:diff2.2.2}
\end{equation}

The characteristic constraint is
\begin{equation}
  -dy \frac{\partial H}{\partial p_x} + dx \frac{\partial H}{\partial p_y} = 0.
  \label{eq:chara2.2}
\end{equation}
Solving within triangle $EGM$ is therefore equivalent to solving \eqref{eq:eikonal}, \eqref{eq:diff2.2.1}, \eqref{eq:diff2.2.2}, and \eqref{eq:chara2.2} for the unknowns $\tau$, $p_x$, $p_y$, and $p_z$. Because \eqref{eq:diff2.2.1}, \eqref{eq:diff2.2.2}, and \eqref{eq:chara2.2} are linear, they allow $p_x$, $p_y$, and $p_z$ to be expressed as linear functions of $\tau$. Substituting these expressions into \eqref{eq:eikonal} again yields a quartic equation, whose solution gives the update for this case. The resulting $\tau$ is then checked against the causality condition. If the condition is not satisfied, we compute the solutions corresponding to the characteristic directions along $\overrightarrow{EM}$ and $\overrightarrow{GM}$ separately and select the smaller valid value.
The local update formula for this case can be expressed as:
\[
\tau_M^{\text{new}} =
\begin{cases}
  \min\left(\tau_M^{\text{old}}, \tau_1^{\text{valid}}, \tau_2^{\text{valid}}\right), & \text{if two valid roots } \tau_1^{\text{valid}}, \tau_2^{\text{valid}} \text{ exist}, \\[6pt]
  \min\left(\tau_M^{\text{old}}, \tau^{\text{valid}}\right), & \text{if only one valid root } \tau^{\text{valid}} \text{ exists}, \\[6pt]
  \min\left(\tau_M^{\text{old}}, \dfrac{T_{0E}\tau_E + \dfrac{|\overrightarrow{EM}|}{v_{\overrightarrow{EM}}}}{T_{0M}}, \dfrac{T_{0G}\tau_G + \dfrac{|\overrightarrow{GM}|}{v_{\overrightarrow{GM}}}}{T_{0M}}\right), & \text{if no valid root exists},
\end{cases}
\]
where $v_{\overrightarrow{EM}}$ and $v_{\overrightarrow{GM}}$ are the group velocities along the characteristic directions
$\overrightarrow{EM}$ and $\overrightarrow{GM}$, respectively. 

The configuration in \Cref{fig:img2.2} is similar to that in \Cref{fig:img2.1}. Only the finite-difference relations change; the characteristic constraint remains the same. Therefore, $p_x$ and $p_y$ satisfy the same proportional relation in both cases.

\paragraph{Case 2: Vertical--Horizontal Mixed Constraint}

For \Cref{fig:img2.3}, the spatial increments are
\[
dx = x_E - x_G, \quad dy = y_M - y_G, \quad dz = z_E - z_G,
\]
so that
\begin{subequations}
  \begin{align}
    \tau_E - \tau_G &= dx \frac{\partial \tau}{\partial x} + dz \frac{\partial \tau}{\partial z}, \\
    \tau_M - \tau_G &= dy \frac{\partial \tau}{\partial y}.
  \end{align}
\end{subequations}

Substituting these relations into \eqref{eq:grad_fac} gives
\begin{equation}
  \left(dx \frac{\partial T_0}{\partial x} + dz \frac{\partial T_0}{\partial z} \right) \tau + T_0 (\tau_E - \tau_G) = dx p_x + dz p_z,
  \label{eq:diff2.3.1}
\end{equation}
\begin{equation}
  dy \frac{\partial T_0}{\partial y} \tau + T_0 (\tau - \tau_G) = dy p_y.
  \label{eq:diff2.3.2}
\end{equation}

The characteristic constraint is
\begin{equation}
  -dx \frac{\partial H}{\partial p_z} + dz \frac{\partial H}{\partial p_x} = 0.
  \label{eq:chara2.3}
\end{equation}

The full system, consisting of \eqref{eq:diff2.3.1}, \eqref{eq:diff2.3.2}, \eqref{eq:eikonal}, and \eqref{eq:chara2.3}, is solved by the bisection method. For a given trial value of $p_z$, we first compute $p_x^2 + p_y^2$ from \eqref{eq:eikonal}. We then evaluate $\frac{\partial H}{\partial p_z}$ and use \eqref{eq:chara2.3} to determine $p_x$. Substituting these quantities into \eqref{eq:diff2.3.1} gives the corresponding value of $\tau$, after which $p_y$ is recovered from \eqref{eq:diff2.3.2}. The variable $p_z$ is updated iteratively until the governing equations are satisfied. The resulting $\tau$ is then checked against the causality condition. If the condition is not satisfied, we compute the solutions corresponding to the characteristic directions along $\overrightarrow{EM}$ and $\overrightarrow{GM}$ separately and select the smaller valid value.
The local update formula for this case can be expressed as:
\[
\tau_M^{\text{new}} =
\begin{cases}
  \min\left(\tau_M^{\text{old}}, \tau^{\text{valid}}\right), & \text{if only one valid root } \tau^{\text{valid}} \text{ exists}, \\[6pt]
  \min\left(\tau_M^{\text{old}}, \dfrac{T_{0E}\tau_E + \dfrac{|\overrightarrow{EM}|}{v_{\overrightarrow{EM}}}}{T_{0M}}, \dfrac{T_{0G}\tau_G + \dfrac{|\overrightarrow{GM}|}{v_{\overrightarrow{GM}}}}{T_{0M}}\right), & \text{if no valid root exists},
\end{cases}
\]
where $v_{\overrightarrow{EM}}$ and $v_{\overrightarrow{GM}}$ are the group velocities along the characteristic directions $\overrightarrow{EM}$ and $\overrightarrow{GM}$, respectively.

The update for \Cref{fig:img2.4} is identical to that for \Cref{fig:img2.3}. Apart from redefining the geometric increments $dx$, $dy$, and $dz$ according to the local configuration, one only needs to interchange $p_x$, $dx$, and $\frac{\partial T_0}{\partial x}$ with $p_y$, $dy$, and $\frac{\partial T_0}{\partial y}$ in \eqref{eq:diff2.3.1}, \eqref{eq:diff2.3.2}, and \eqref{eq:chara2.3}.

\paragraph{Case 3: Alternative Mixed Constraint}
For \Cref{fig:img2.5}, the spatial parameters are
\[
dx = x_M - x_G, \quad dy = y_E - y_G, \quad dz = z_M - z_G,
\]
so that
\begin{subequations}
  \begin{align}
    \tau_E - \tau_G &= dy \frac{\partial \tau}{\partial y}, \\
    \tau_M - \tau_G &=  dx \frac{\partial \tau}{\partial x} + dz \frac{\partial \tau}{\partial z}.
  \end{align}
\end{subequations}
Substituting these relations into \eqref{eq:grad_fac} gives
\begin{equation}
  dy \frac{\partial T_0}{\partial y} \tau + T_0 (\tau_E - \tau_G) = dy p_y,
  \label{eq:diff2.4.1}
\end{equation}
\begin{equation}
  \left(dx \frac{\partial T_0}{\partial x} + dz \frac{\partial T_0}{\partial z} \right) \tau + T_0 (\tau - \tau_G) = dx p_x + dz p_z.
  \label{eq:diff2.4.2}
\end{equation}

The characteristic constraint is the same as in \Cref{fig:img2.3}. The full system consists of \eqref{eq:diff2.4.1}, \eqref{eq:diff2.4.2}, \eqref{eq:eikonal}, and \eqref{eq:chara2.3}, and is again solved by the bisection method. For a given trial value of $p_z$, we compute $p_x^2 + p_y^2$ from \eqref{eq:eikonal}, evaluate $\frac{\partial H}{\partial p_z}$, and then obtain $p_x$ from \eqref{eq:chara2.3}. Substituting these quantities into \eqref{eq:diff2.4.2} gives the corresponding value of $\tau$, and $p_y$ is then recovered from \eqref{eq:diff2.4.1}. The variable $p_z$ is adjusted iteratively until the governing equations are satisfied. The resulting $\tau$ is then checked against the causality condition. If the condition is not satisfied, we compute the solutions corresponding to the characteristic directions along $\overrightarrow{EM}$ and $\overrightarrow{GM}$ separately and select the smallest valid value.
The local update formula for this case can be expressed as:
\[
\tau_M^{\text{new}} =
\begin{cases}
  \min\left(\tau_M^{\text{old}}, \tau^{\text{valid}}\right), & \text{if only one valid root } \tau^{\text{valid}} \text{ exists}, \\[6pt]
  \min\left(\tau_M^{\text{old}}, \dfrac{T_{0E}\tau_E + \dfrac{|\overrightarrow{EM}|}{v_{\overrightarrow{EM}}}}{T_{0M}}, \dfrac{T_{0G}\tau_G + \dfrac{|\overrightarrow{GM}|}{v_{\overrightarrow{GM}}}}{T_{0M}}\right), & \text{if no valid root exists},
\end{cases}
\]
where $v_{\overrightarrow{EM}}$ and $v_{\overrightarrow{GM}}$ are the group velocities along the characteristic directions $\overrightarrow{EM}$ and $\overrightarrow{GM}$, respectively.

The update for \Cref{fig:img2.6} is identical to that for \Cref{fig:img2.5}. Apart from redefining the geometric increments $dx$, $dy$, and $dz$ according to the local configuration, one only needs to interchange $p_x$, $dx$, and $\frac{\partial T_0}{\partial x}$ with $p_y$, $dy$, and $\frac{\partial T_0}{\partial y}$ in \eqref{eq:diff2.4.1}, \eqref{eq:diff2.4.2}, and \eqref{eq:chara2.3}.

\paragraph{Existence and Uniqueness of the Bisection Solution}

Taking Case 2 as an example, suppose that $|p_z|$ increases while its sign is fixed by causality. Then \eqref{eq:eikonal} implies that $p_x^2+p_y^2$ decreases. By \eqref{eq:dH_dpz}, this makes $\left|\frac{\partial H}{\partial p_z}\right|$ increase, and \eqref{eq:chara2.3} then implies that $|p_x|$ also increases. The corresponding $\tau$ obtained from \eqref{eq:diff2.3.1} therefore increases, and so does $|p_y|$ through \eqref{eq:diff2.3.2}. Hence $|p_x|$, $|p_y|$, and $|p_z|$ increase simultaneously, so the left-hand side of \eqref{eq:eikonal} varies monotonically while the right-hand side remains fixed. Existence can therefore be checked by verifying that the residual changes sign across the admissible interval for $p_z$, and uniqueness follows from the same monotonicity.

\subsubsection{Algorithm}

The overall algorithm is summarized in Algorithm~\ref{alg:fsm}. The reference field $T_0$ is precomputed once using the analytical solution in a homogeneous medium with parameters taken from the source location. The perturbation factor $\tau$ is initialized to $1$ at the eight grid points surrounding the source and to a large value elsewhere. Sweeps are performed in eight alternating directions until the maximum change in $\tau$ falls below a tolerance (here $10^{-4}$). At each grid point, the update is computed for the tetrahedron that corresponds to the current sweep direction, using the appropriate formula (tetrahedral interior or face). The smallest admissible value among all causal candidates is accepted.

\begin{algorithm}
  \caption{FSM for 3D  VTI factored eikonal equation using six-tetrahedron pyramidal stencil}
  \label{alg:fsm}
  \begin{algorithmic}[1]
    \State Compute $T_0$ analytically using source‑location parameters.
    \State Initialize $\tau = 1$ at the eight grid points around the source; set $\tau = \infty$ elsewhere.
    \While{not converged}
    \For{each sweep direction $d = 1,\dots,8$}
    \For{each grid point $M$ in the order defined by $d$}
    \State Determine the tetrahedron (or face) for the current sweep direction.
    \If{tetrahedral interior update is applicable}
    \State Form the quartic coefficients and solve for $\tau$.
    \If{a causal root is found}
    \State Tentative $\tau_{\text{new}} \gets$ root value.
    \Else
    \State Evaluate three face updates and take the smallest causal value.
    \EndIf
    \ElsIf{inclined face update is applicable}
    \If{horizontally constrained case}
    \State Form quartic from linearized $p_x,p_y$ and solve.
    \ElsIf{mixed constrained case}
    \State Perform bisection on $p_z$ (or equivalent) to find $\tau$.
    \EndIf
    \If{no causal root found}
    \State Fall back to one‑segment updates along the two edges.
    \EndIf
    \Else
    \State Use axis‑aligned face updates (direct quadratic solves).
    \EndIf
    \State $\tau_M \gets \min(\tau_M, \tau_{\text{new}})$.
    \EndFor
    \EndFor
    \State Check convergence by $\max|\tau_{\text{new}} - \tau_{\text{old}}|$.
    \EndWhile
    \State Recover $T = T_0 \tau$.
  \end{algorithmic}
\end{algorithm}

\section{Numerical Examples}

\subsection{Homogeneous model}

We first consider a homogeneous 3D VTI model with $V_{\text{nmo}} = 2500\ \text{m/s}$, $V_{0} = 2000\ \text{m/s}$, and $\eta = 0.2$. The model size is $101\times 101\times 51$, with grid spacings $dx=dy=dz=100$ m, and the source is placed at $(x_{s},y_{s},z_{s}) = (6000,5000,2000)$ m. We compute traveltimes using both the direct finite-difference (FD) method and the proposed factored solver, and compare them with a reference solution obtained by the shooting method. 

\begin{figure}[!htbp]
  \centering
  \subfloat[Direct FD]{\includegraphics[width=0.5\linewidth]{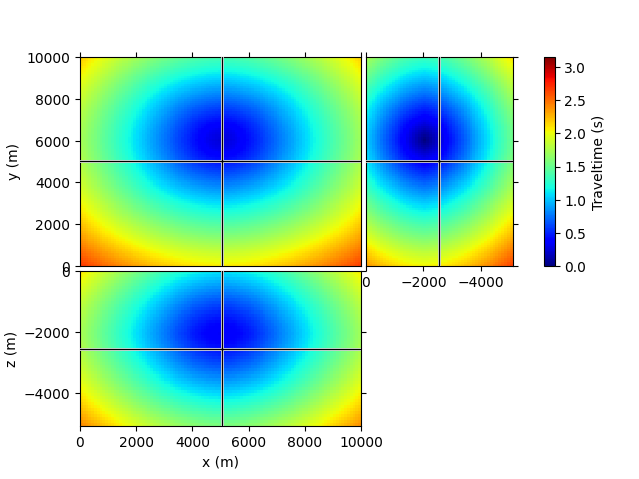}\label{fig:direct}}
  \subfloat[Factorization]{\includegraphics[width=0.5\linewidth]{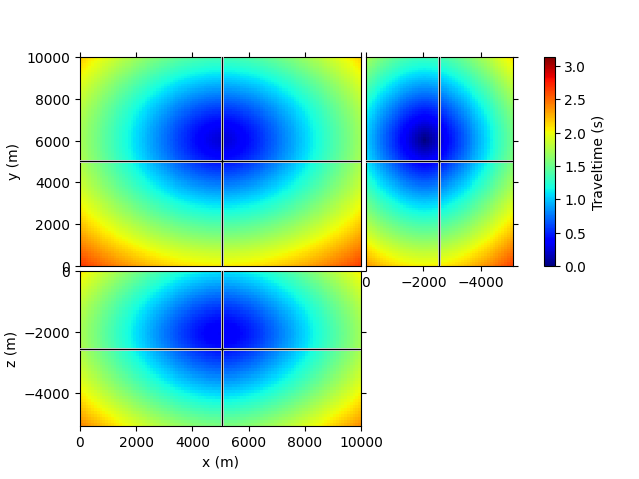}\label{fig:factorization}}\\
  \subfloat[Error of direct FD]{\includegraphics[width=0.5\linewidth]{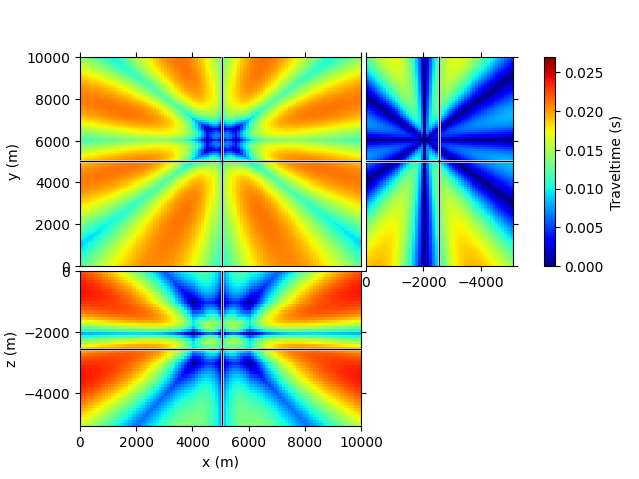}\label{fig:directerror}}
  \subfloat[Error of factorization]{\includegraphics[width=0.5\linewidth]{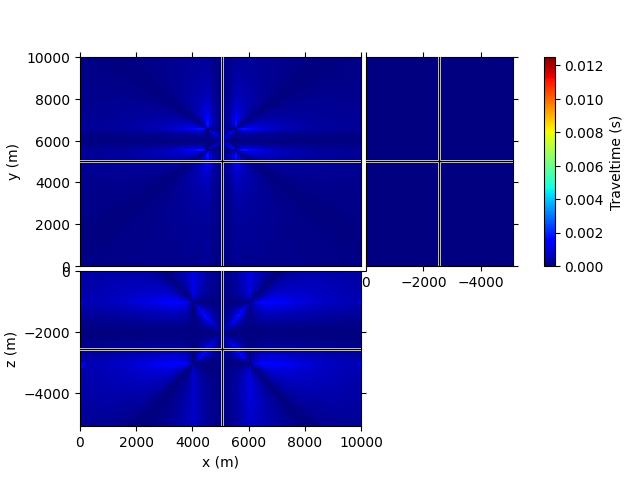}\label{fig:factorizationerror}}\\
  \caption{Traveltimes computed by direct FD and factorization, together with the errors relative to the reference solution obtained by the shooting method.}\label{fig:traveltime3d}
\end{figure}

\begin{figure}[!htbp]
  \centering
  \subfloat[]{\includegraphics[width=0.32\linewidth]{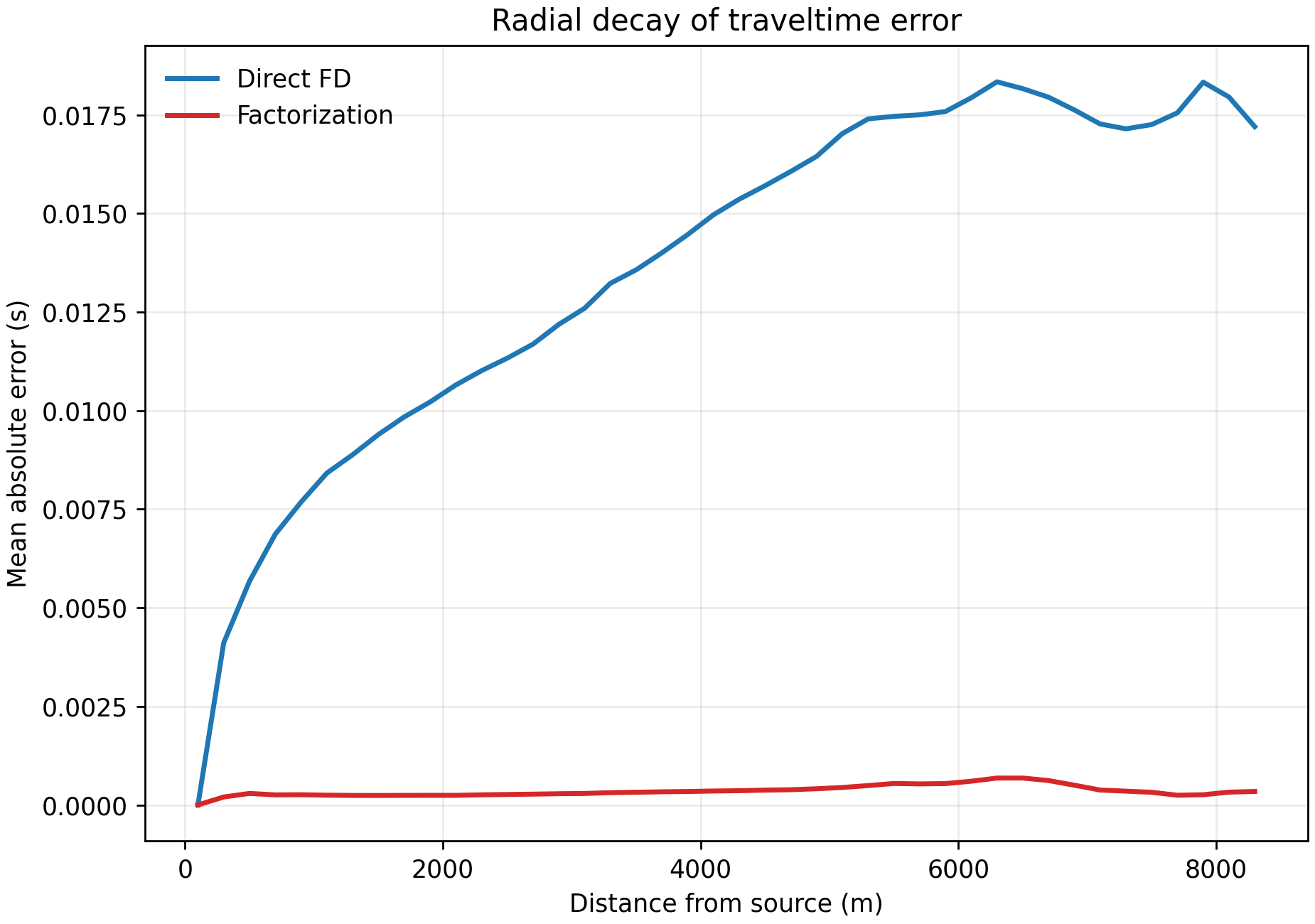}\label{fig:homogeneous:radial}}
  \subfloat[]{\includegraphics[width=0.32\linewidth]{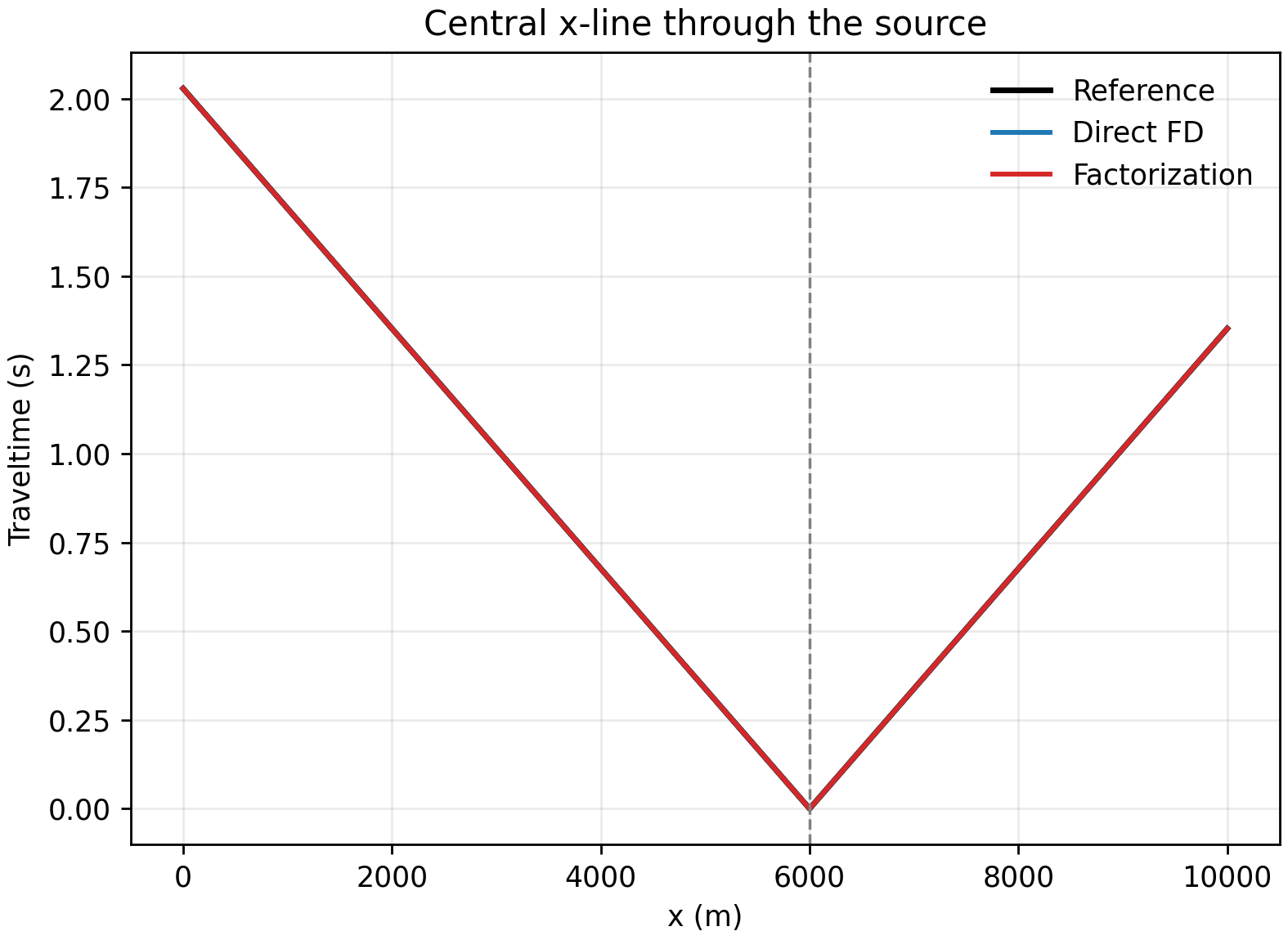}\label{fig:homogeneous:line:a}}
  \subfloat[]{\includegraphics[width=0.32\linewidth]{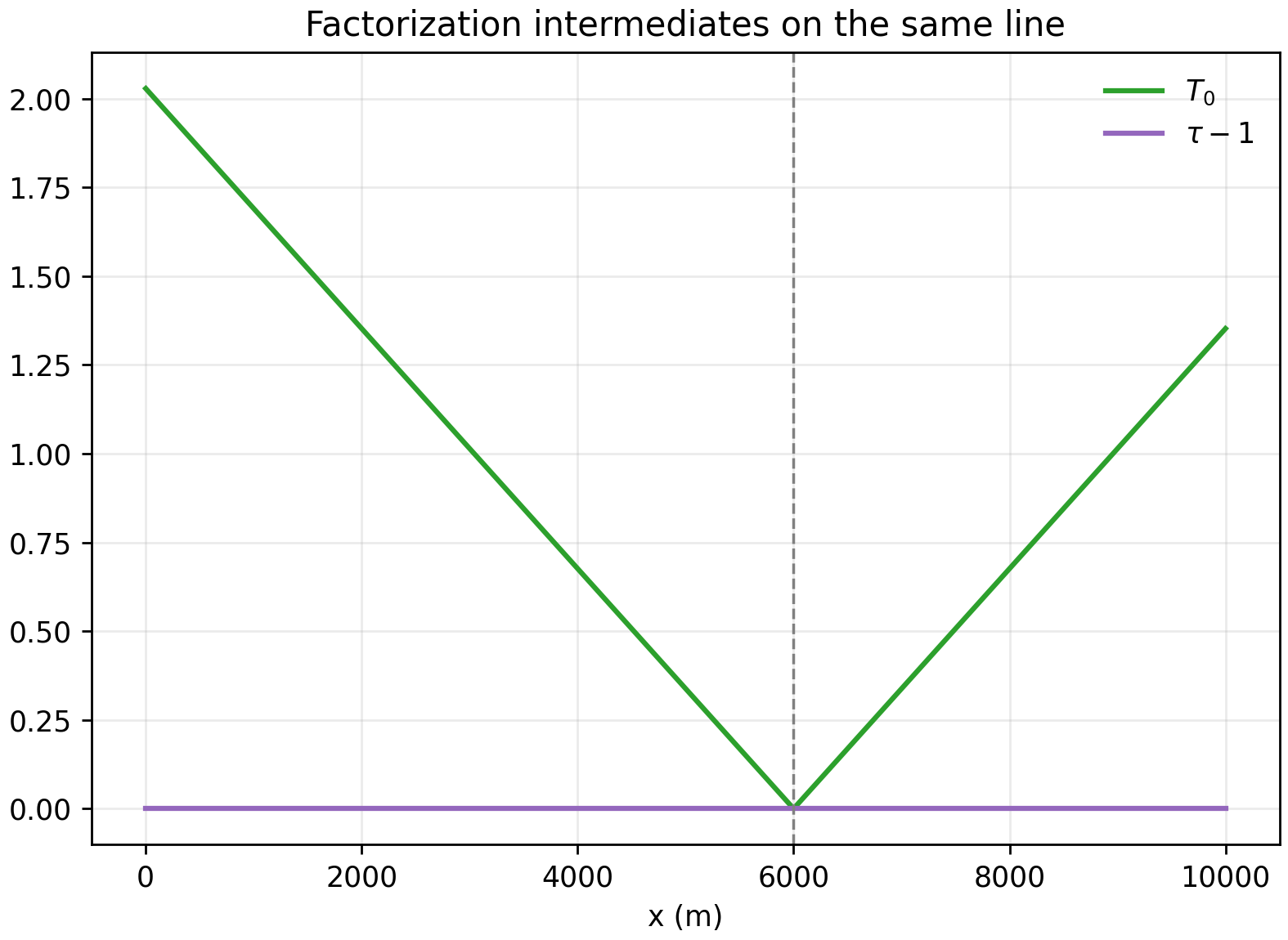}\label{fig:homogeneous:line:b}}
  \caption{(a) Mean absolute error decays away from the source. (b) Traveltime profiles on the central $x$ line of the reference, direct-FD, and factorized traveltimes. (c) The corresponding factorization variables $T_0$ and $\tau-1$ on the same line.}\label{fig:homogeneous:diagnostics}
\end{figure}

The computed traveltimes and their errors are shown in \Cref{fig:traveltime3d}. Both methods recover the overall traveltime pattern, but the direct FD solution exhibits substantially larger errors, especially near the source, whereas the factored solver produces a much cleaner error field. This behavior is consistent with the role of multiplicative factorization, which regularizes the source neighborhood and improves the quality of the local finite-difference update. The diagnostics in \Cref{fig:homogeneous:diagnostics} further clarify this improvement: the reduction in error is strongest near the source, but remains visible at larger distances as well; along the central line the factorized traveltime is nearly indistinguishable from the reference solution, while \Cref{fig:homogeneous:line:b} displays the corresponding behavior of the factorization variables.

\begin{figure}[!htbp]
  \centering
  \subfloat[Slices of $\tau-1$]{\includegraphics[width=0.32\linewidth]{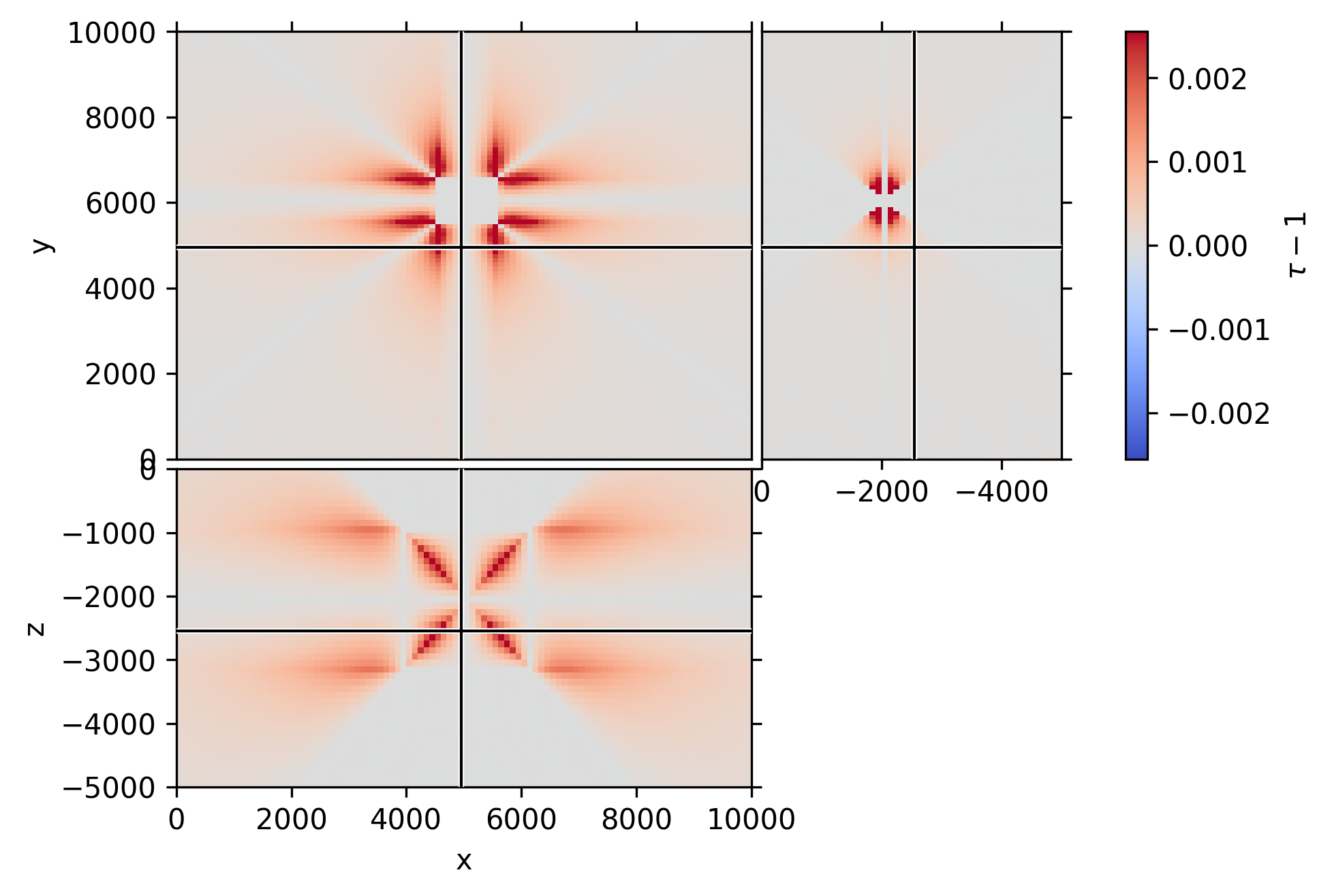}\label{fig:homogeneous:correction:a}}
  \subfloat[Radial decay of $|\tau-1|$]{\includegraphics[width=0.32\linewidth]{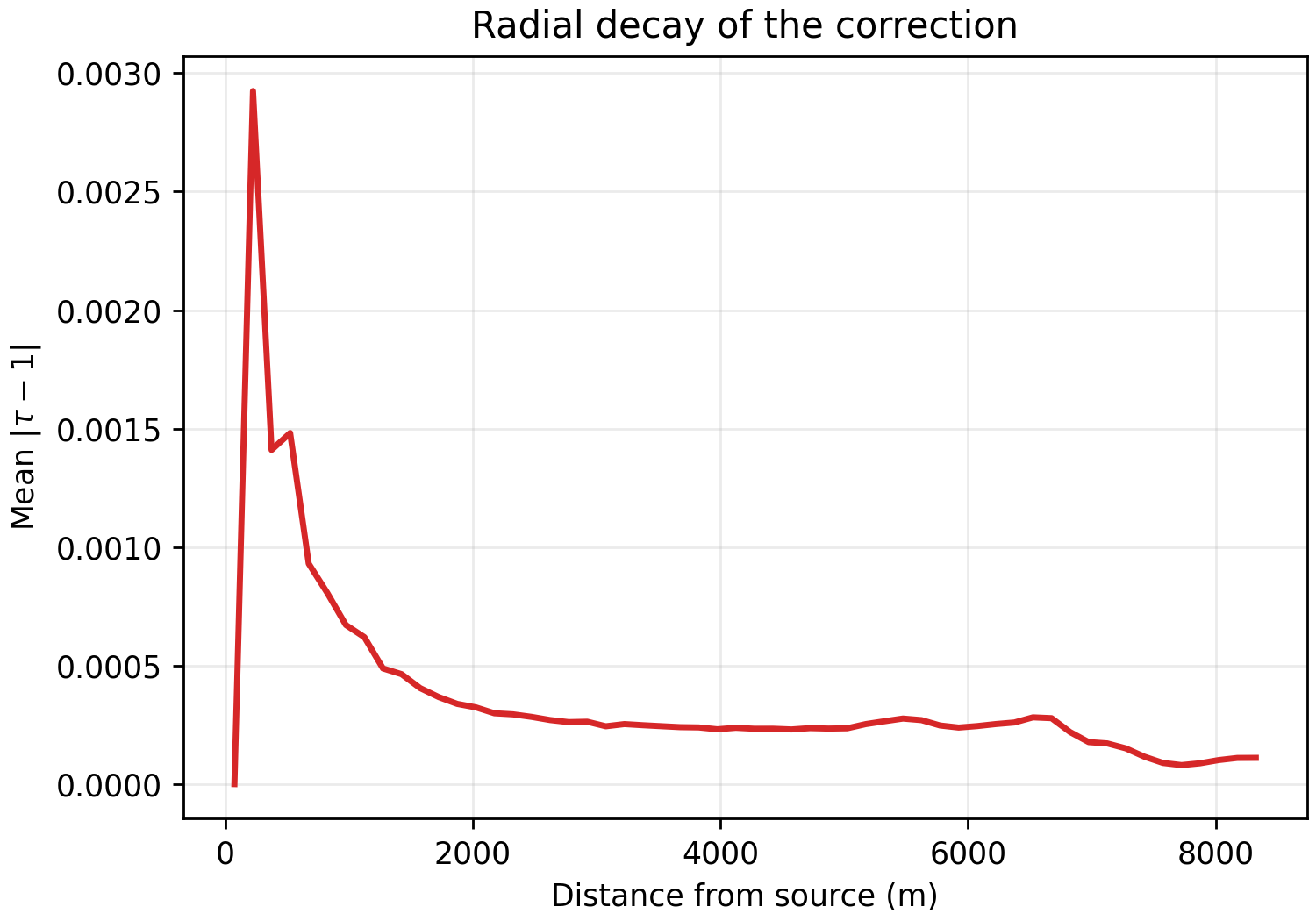}\label{fig:homogeneous:correction:b}}
  \subfloat[Distribution of $|\tau-1|$]{\includegraphics[width=0.32\linewidth]{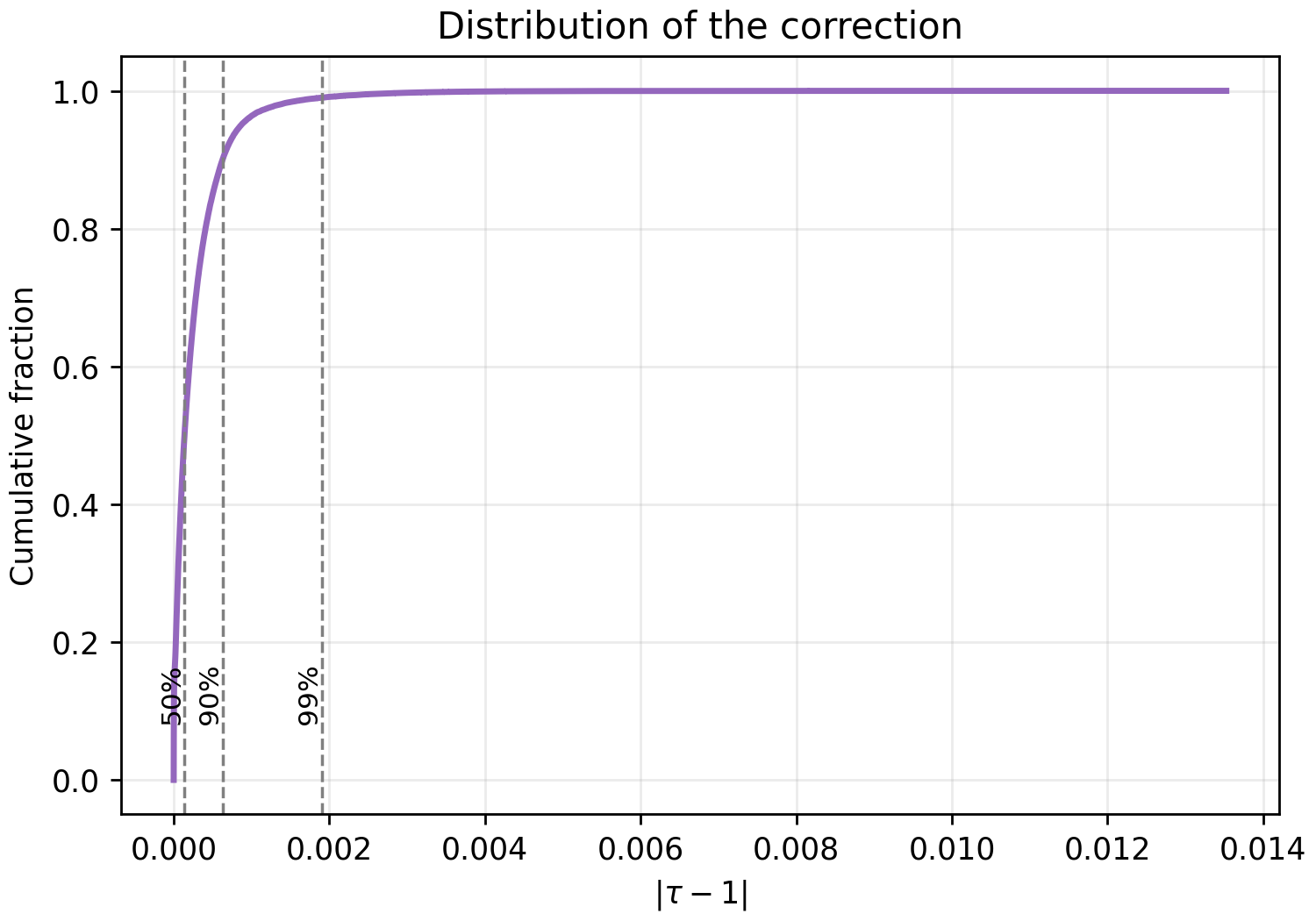}\label{fig:homogeneous:correction:c}}
  \caption{Regularity of the factorized unknown in the homogeneous example. The correction field remains small and smooth after the singular reference field $T_0$ has been removed.}\label{fig:homogeneous:correction}
\end{figure}

\Cref{fig:homogeneous:correction} complements this error analysis by focusing on the unknown solved by the factorized method rather than on the final traveltime itself. Once the singular reference field $T_0$ is removed, the correction field shows no singular spike near the source and remains small over most of the domain. In particular, the median, 90th-percentile, and 99th-percentile values of $|\tau-1|$ are $1.39\times 10^{-4}$, $6.28\times 10^{-4}$, and $1.90\times 10^{-3}$, respectively, while $\max |\tau-1| = 1.35\times 10^{-2}$. These values confirm that the factorized solver does not reconstruct the full singular traveltime directly; instead, it updates a bounded and much smoother correction on top of the analytically motivated reference field.

\begin{figure}[!htbp]
  \centering
  \subfloat[$a$]{\includegraphics[width=0.32\linewidth]{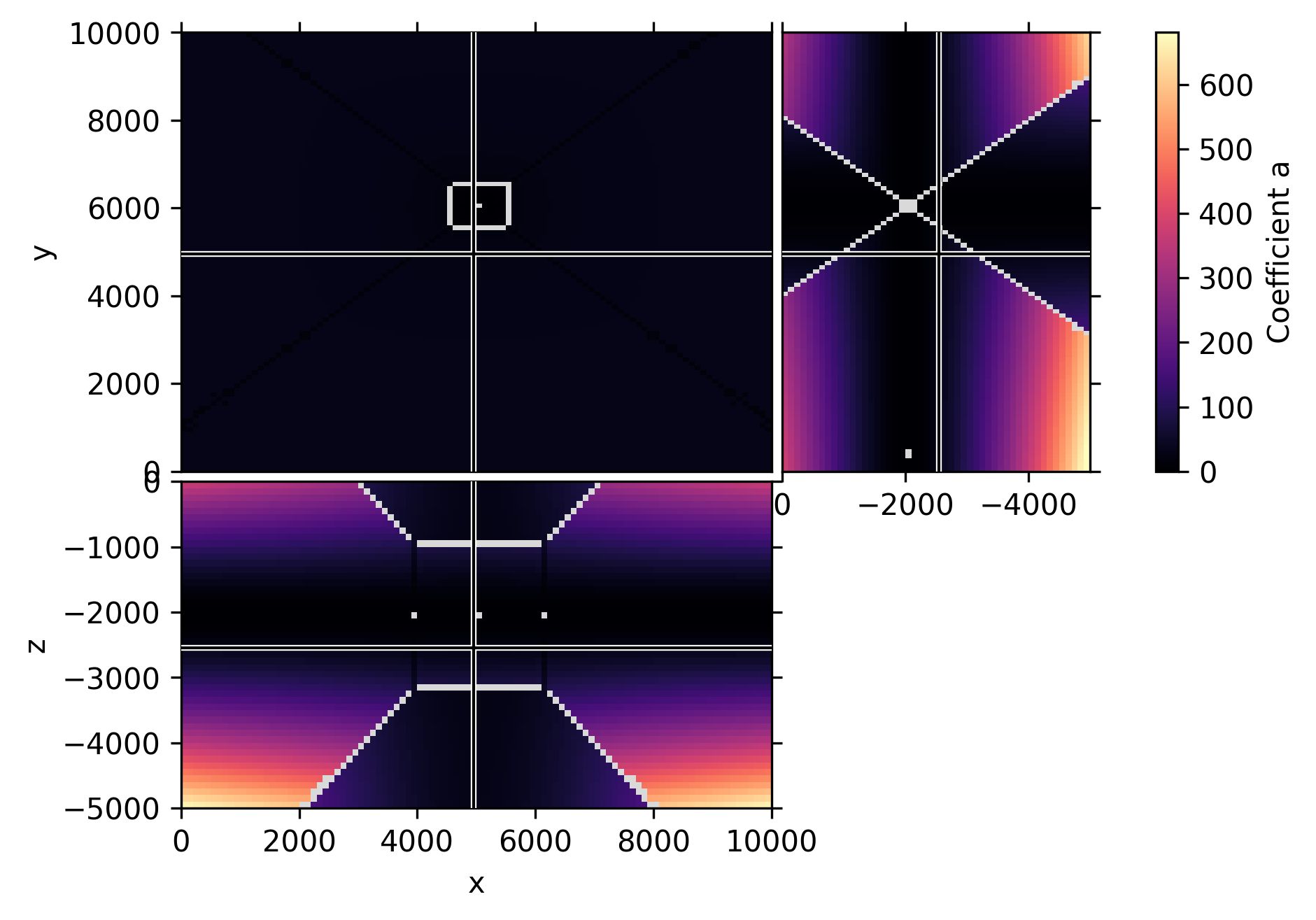}}
  \subfloat[$b$]{\includegraphics[width=0.32\linewidth]{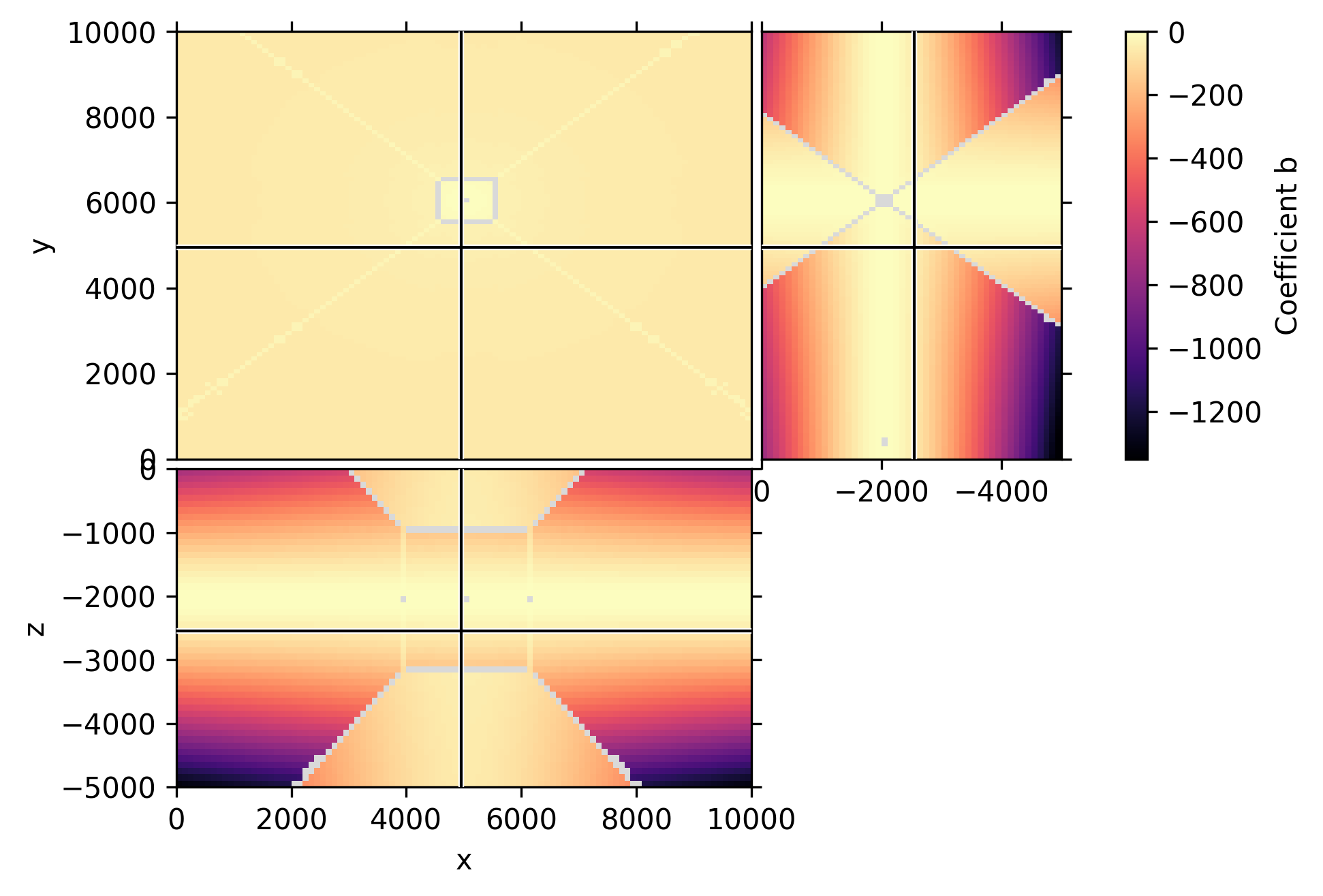}}
  \subfloat[$c$]{\includegraphics[width=0.32\linewidth]{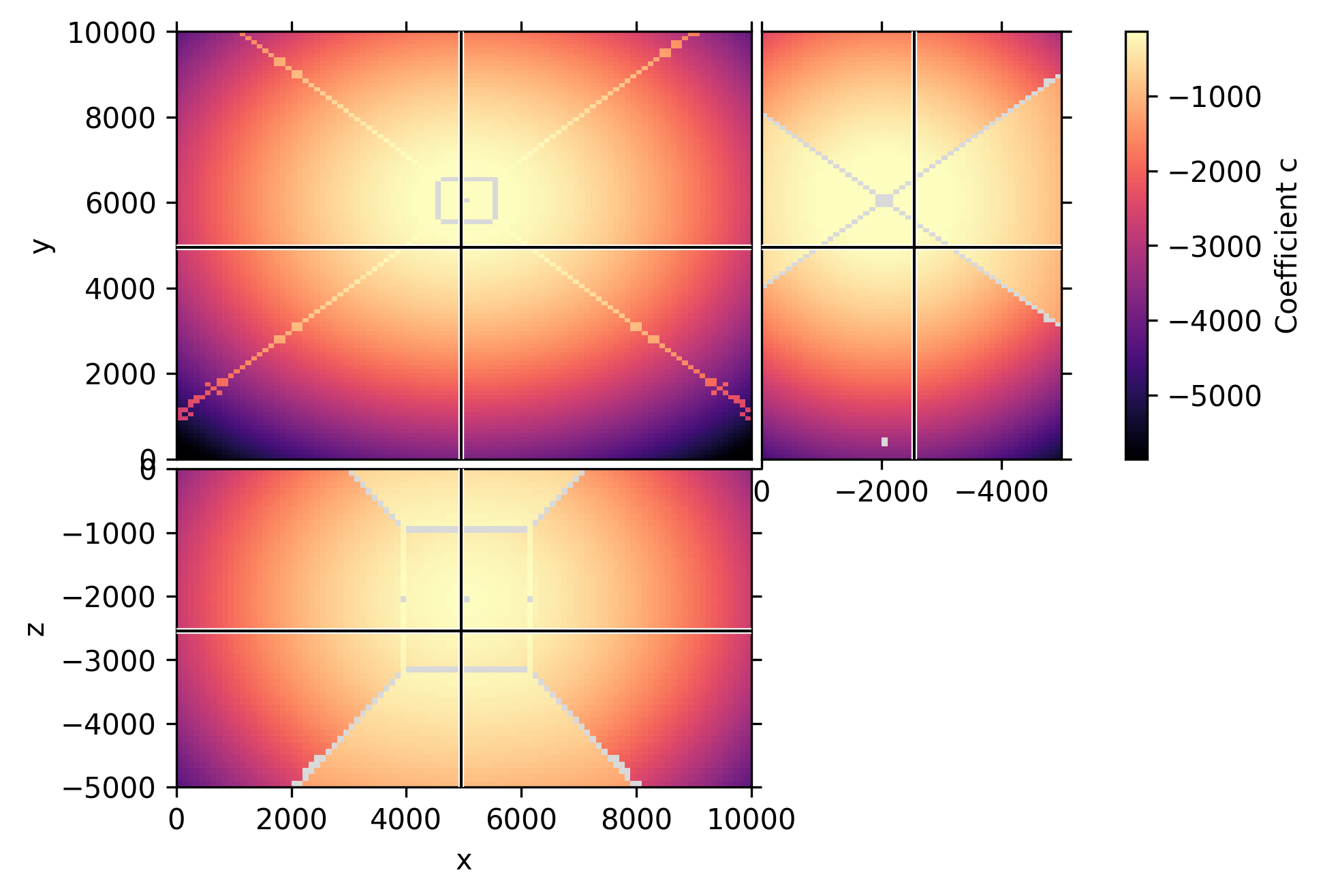}}\\
  \subfloat[$d$]{\includegraphics[width=0.32\linewidth]{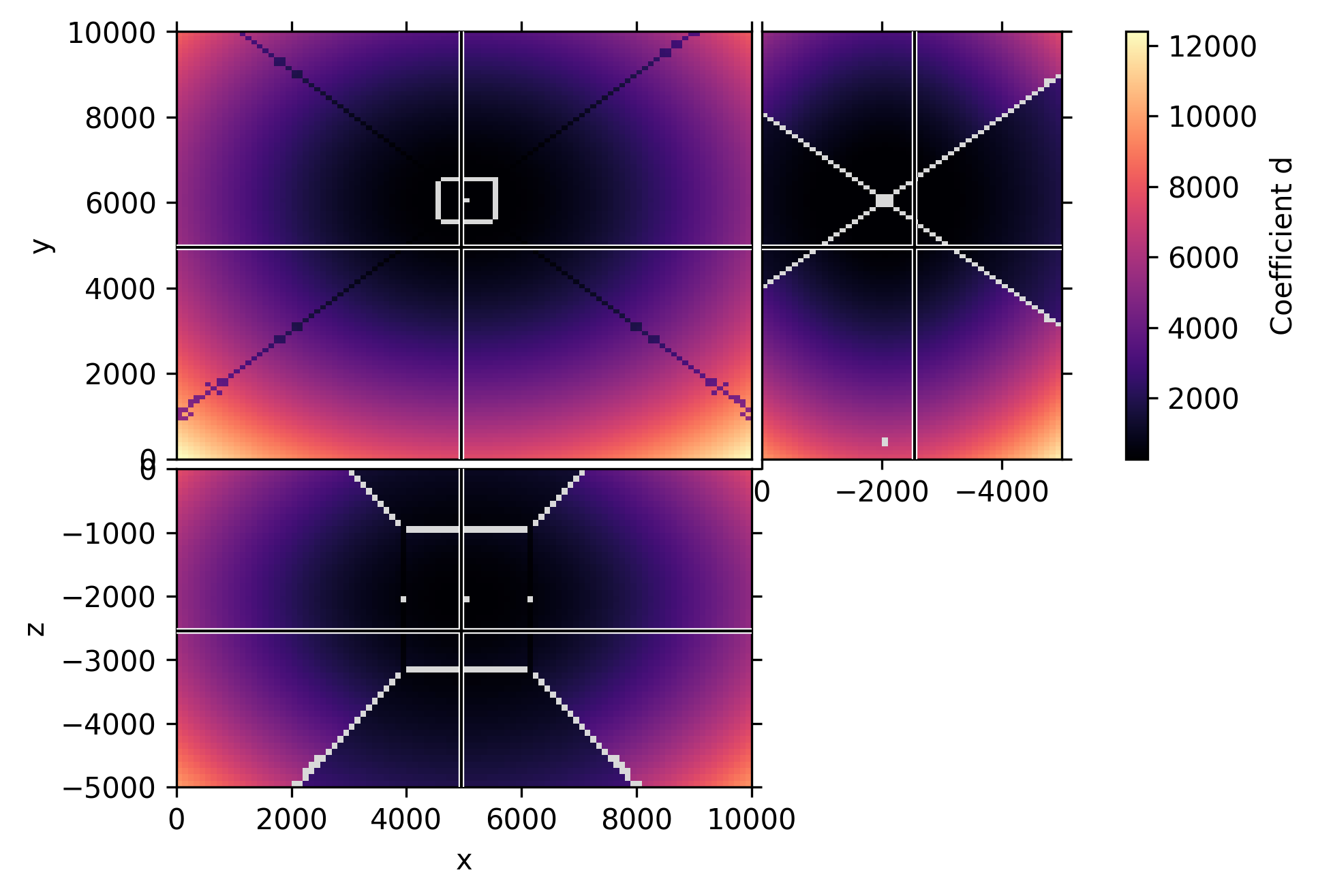}}
  \subfloat[$e$]{\includegraphics[width=0.32\linewidth]{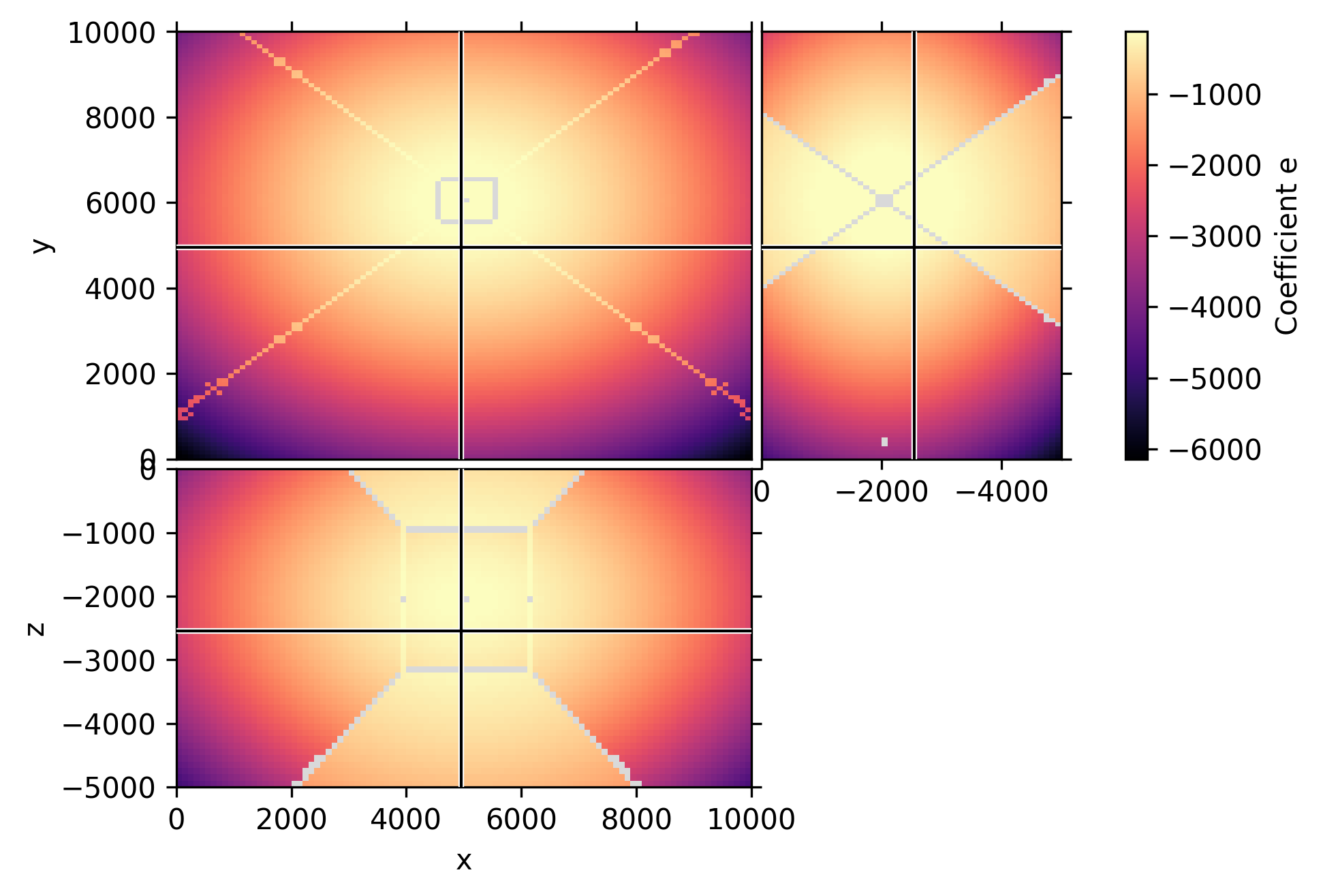}}
  \caption{The coefficients of the local quartic equation \eqref{eq:quartic} in the homogeneous model. The coefficient fields are shown only at grid points where the accepted update came from a quartic solve; cells updated by one-face or mixed-constraint fallback branches are rendered in gray.}\label{fig:homogeneous:quartic}
\end{figure}

\Cref{fig:homogeneous:quartic} makes the algebraic structure of the factorized update explicit by plotting the accepted coefficients $a$, $b$, $c$, $d$, and $e$ of \eqref{eq:quartic}. These coefficient panels should be interpreted as branch-dependent diagnostics rather than as primary solution variables. At each grid point, the solver evaluates several admissible local updates and accepts the smallest causal one. Only some of these candidates come from solving the quartic equation \eqref{eq:quartic}; other accepted updates come from face-based or mixed-constraint fallback branches. Accordingly, the coefficients $a$--$e$ are stored only at grid points where the winning update is quartic, whereas fallback-update cells are rendered in gray because no accepted quartic is associated with them. The resulting support of the coefficient fields is therefore not the whole domain, but the subset of cells where the quartic branch is actually active.
In \Cref{fig:homogeneous:quartic}, the coefficient support is concentrated in the region where the full tetrahedral quartic update is selected by the causality test, whereas other regions are handled more economically by lower-dimensional fallback updates. In this homogeneous test, $517{,}482$ out of $520{,}251$ grid points select a quartic root as the accepted update. Over those quartic-update cells, the coefficient ranges are $a\in[0,\,8.00\times 10^{2}]$, $b\in[-1.58\times 10^{3},\,0]$, $c\in[-7.92\times 10^{3},\,-7.55]$, $d\in[1.08\times 10^{1},\,1.73\times 10^{4}]$, and $e\in[-8.59\times 10^{3},\,-2.85]$. These values confirm that  the quartic coefficients remain bounded and smooth, where the quartic branch is actually active.

The quantitative comparison in \Cref{table:performance} is consistent with these observations. Relative to direct FD, the proposed method reduces the maximum absolute error from $0.026971$ s to $0.011613$ s and the average absolute error from $0.014873$ s to $0.000393$ s. Within a radius of 1.8 km from the source, the mean absolute error decreases from $8.89\times 10^{-3}$ s to $2.46\times 10^{-4}$ s. The factorized solver therefore delivers a much more accurate traveltime field at only a moderate increase in runtime.

\begin{table}[htbp]
  \centering
  \caption{Runtime and traveltime errors for direct FD and factorization.}\label{table:performance}
  \begin{tabular}{l|ccc}

\hline
Method & Runtime (s) & Maximum absolute error (s) & Average absolute error (s) \\
\hline
Factorization & 36.7 & 0.012494 & 0.000447 \\
Direct FD & 26.8 & 0.026971 & 0.014873 \\
\hline
  \end{tabular}
\end{table}

\subsection{Overthrust model}

\begin{figure}[!htbp]
  \centering
  \subfloat[]{\includegraphics[width=0.32\linewidth]{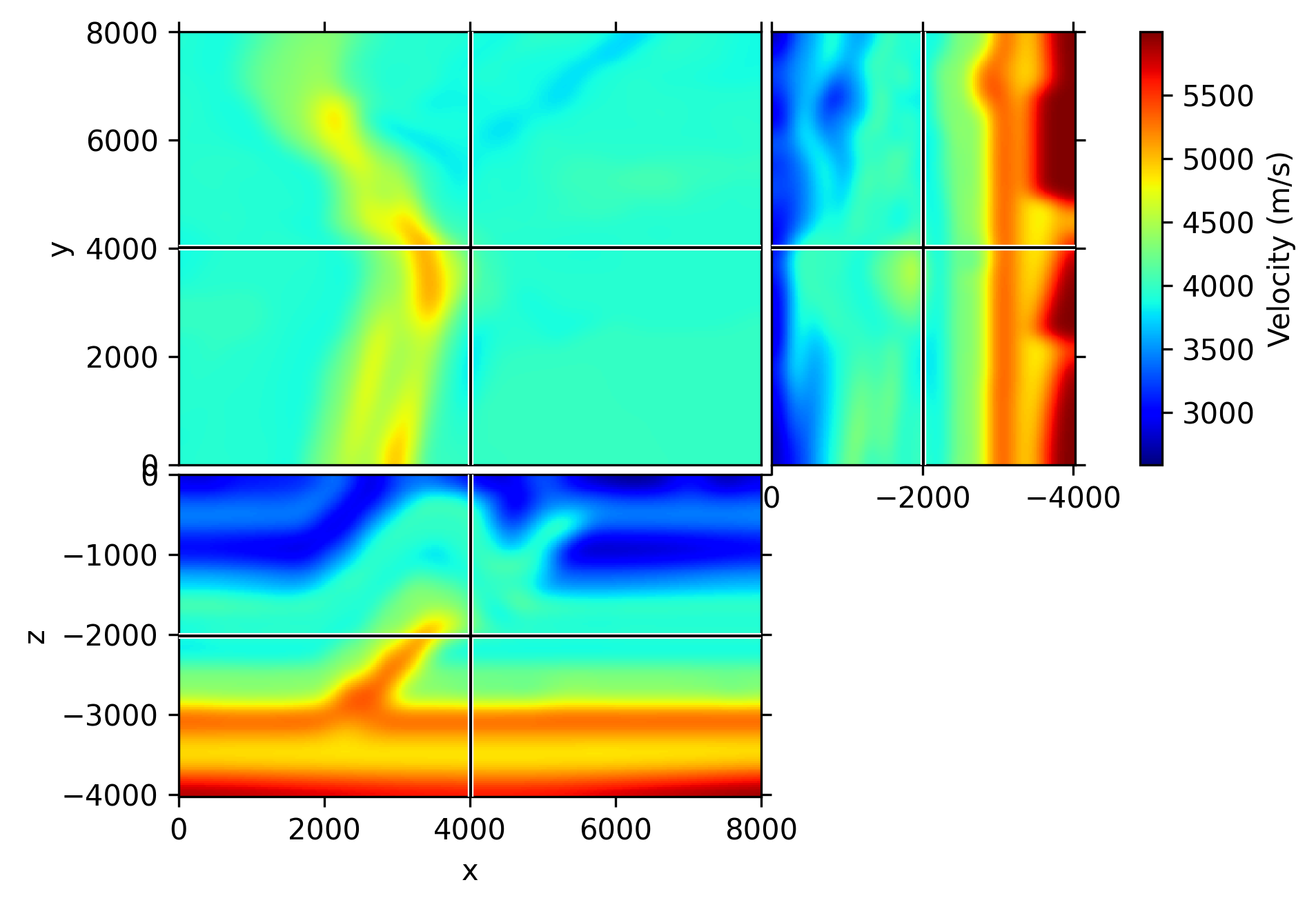}}
  \subfloat[]{\includegraphics[width=0.32\linewidth]{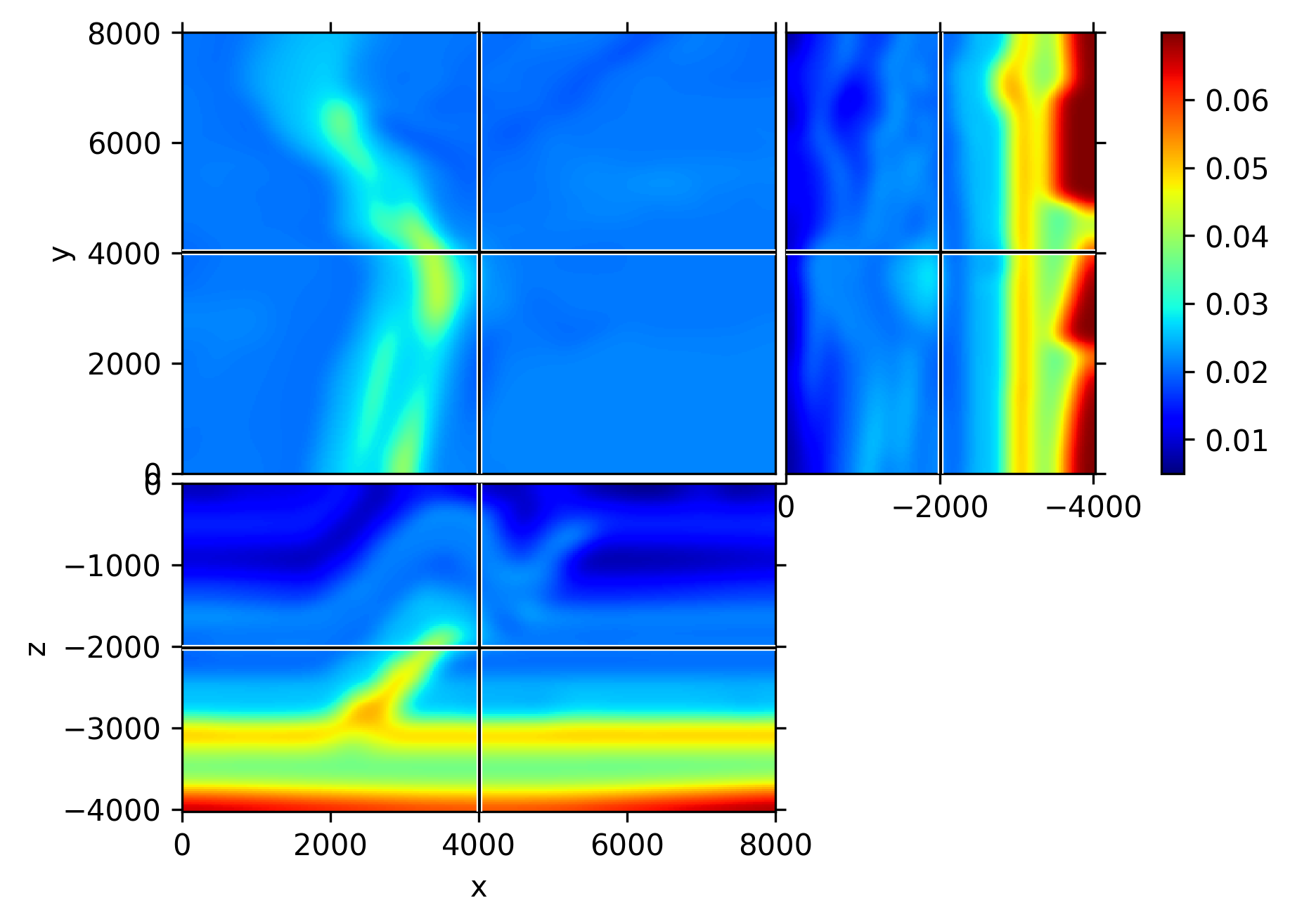}}
  \subfloat[]{\includegraphics[width=0.32\linewidth]{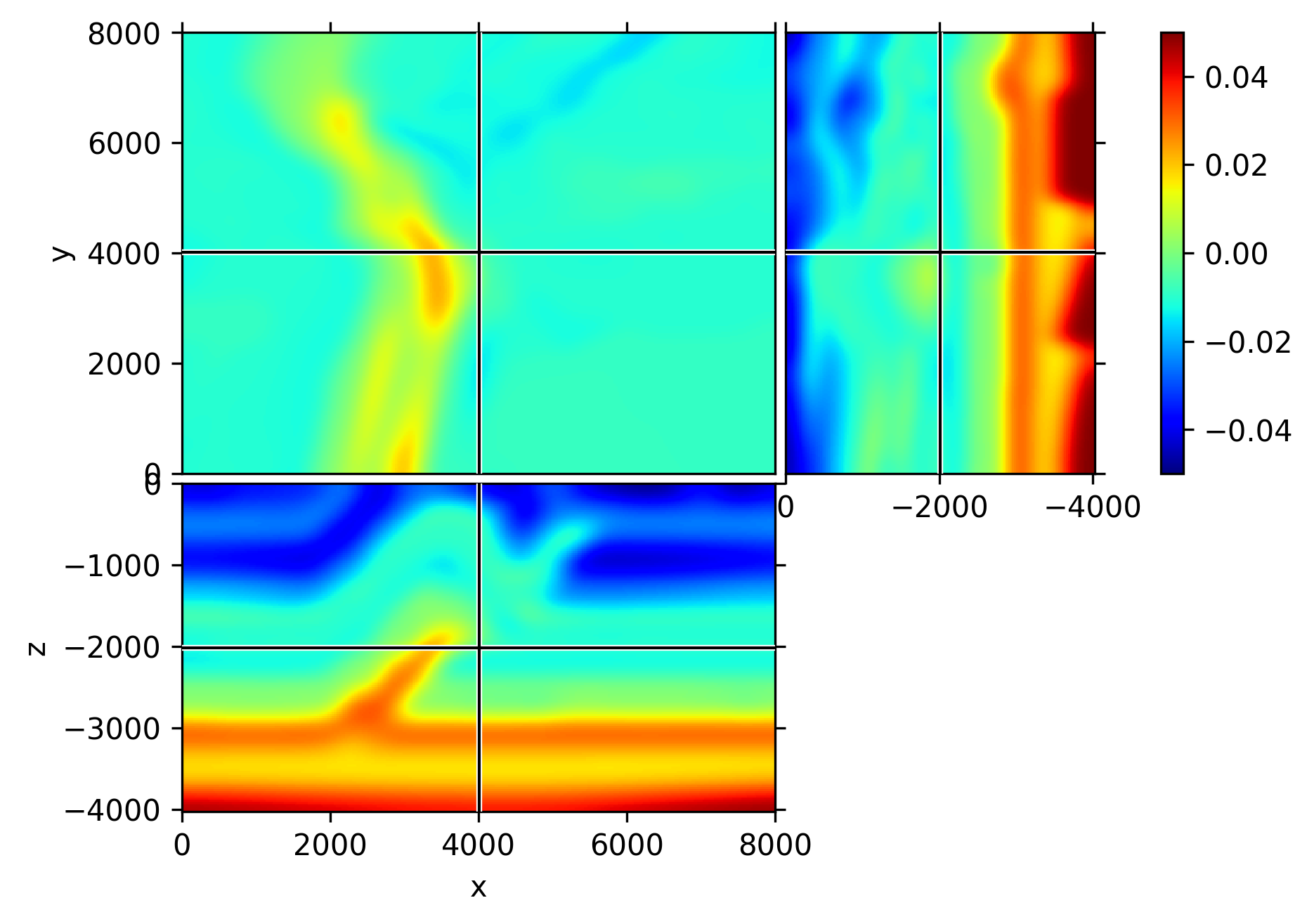}}
  \caption{Overthrust model parameters: (a) $V_p$, (b) $\epsilon$ and (c) $\delta$.}\label{fig:over:params}
\end{figure}

\begin{figure}[htbp]
  \centering
  \subfloat[]{\includegraphics[width=0.32\linewidth]{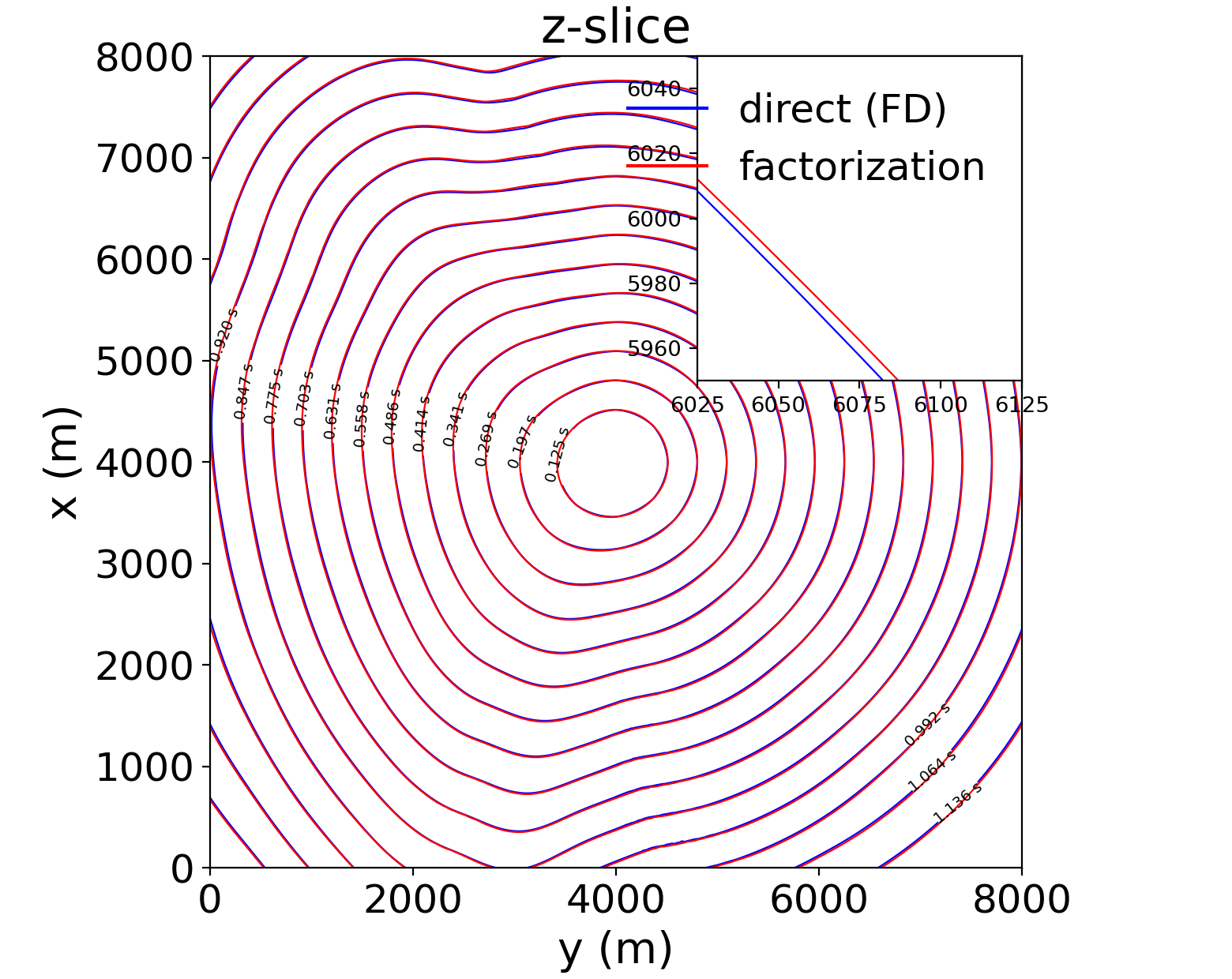}}
  \subfloat[]{\includegraphics[width=0.32\linewidth]{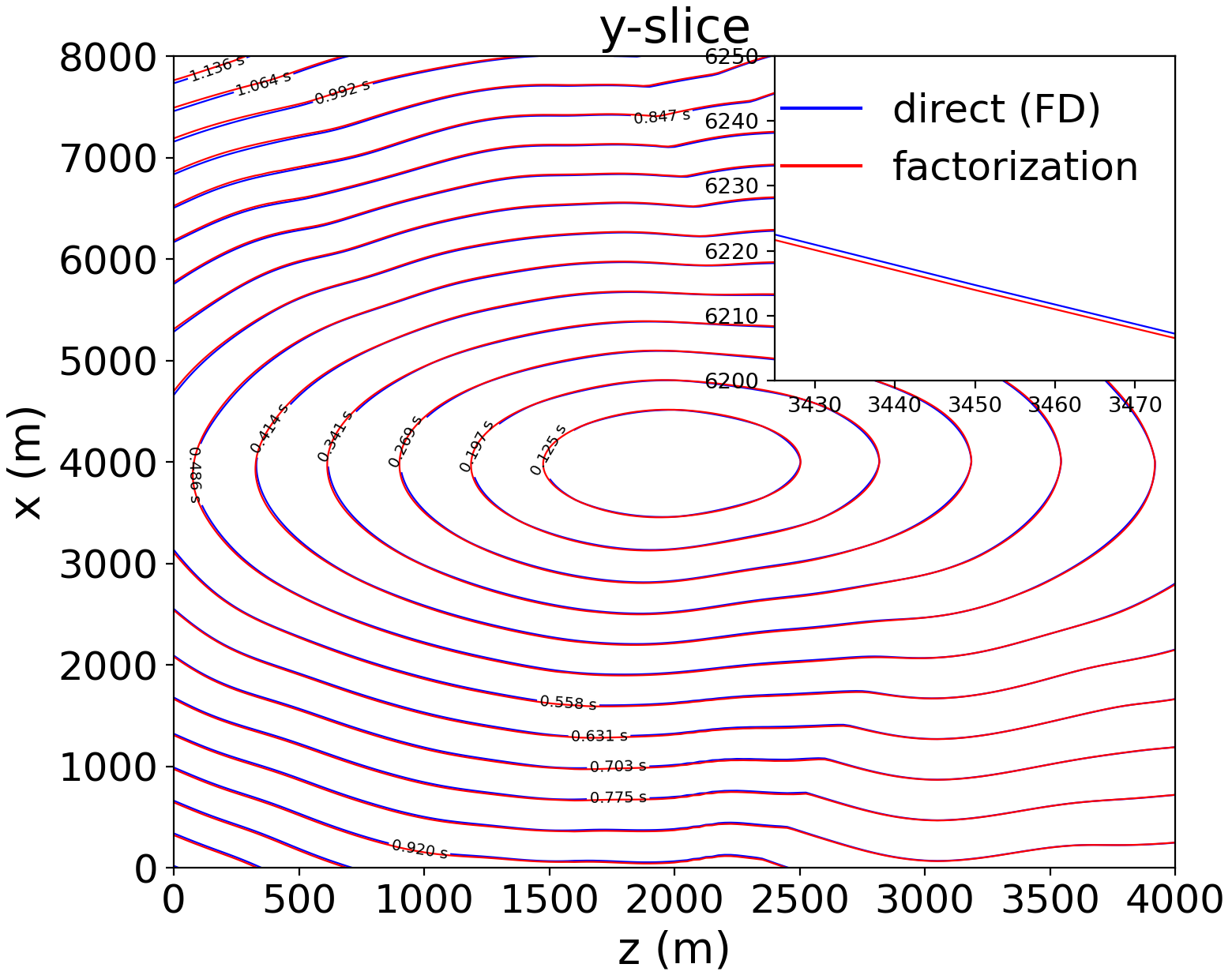}}
  \subfloat[]{\includegraphics[width=0.32\linewidth]{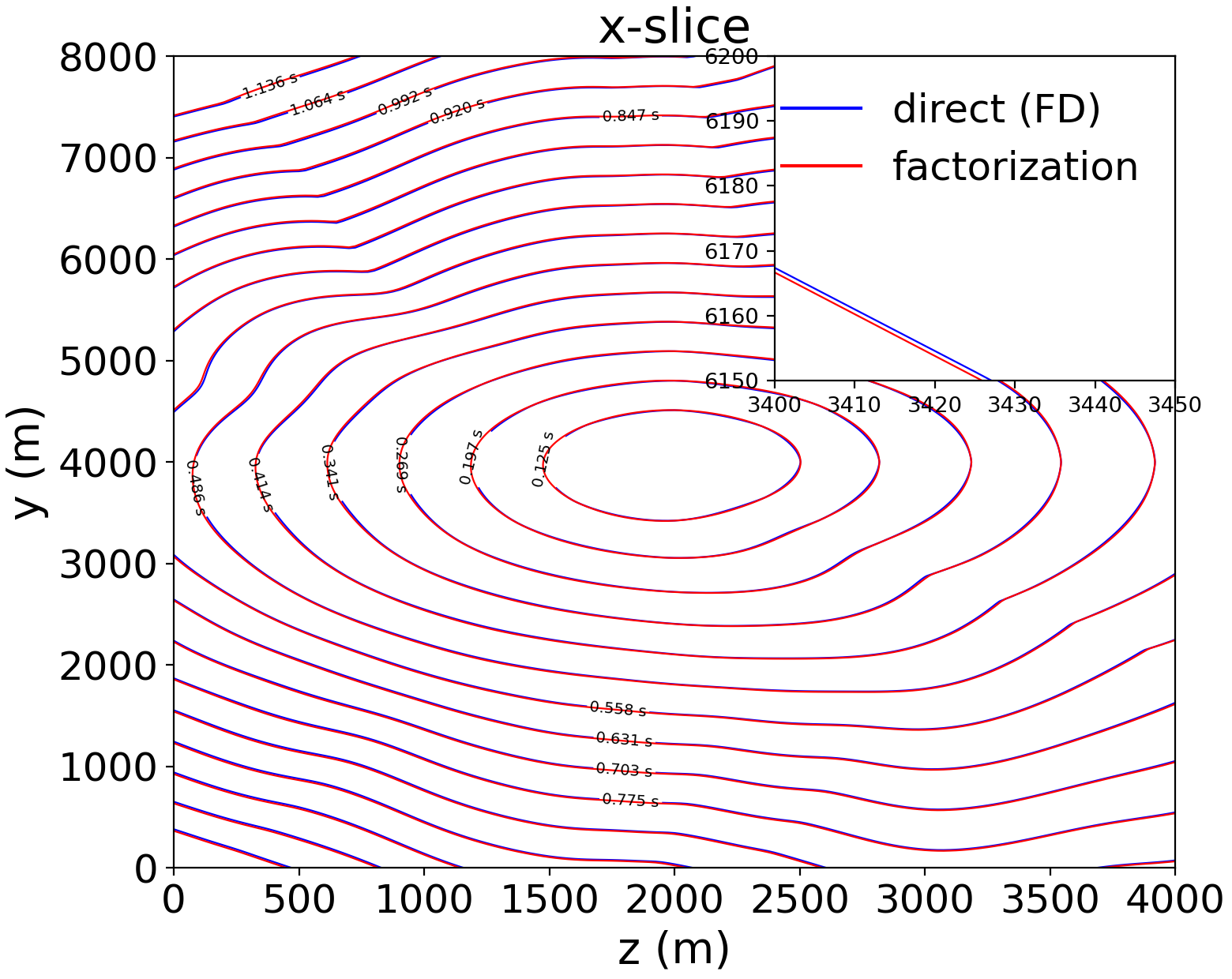}}
  \caption{Traveltime fields computed in the Overthrust model.}\label{fig:over:tt}
\end{figure}

\begin{figure}[htbp]
  \centering
  \subfloat[]{\includegraphics[width=0.32\linewidth]{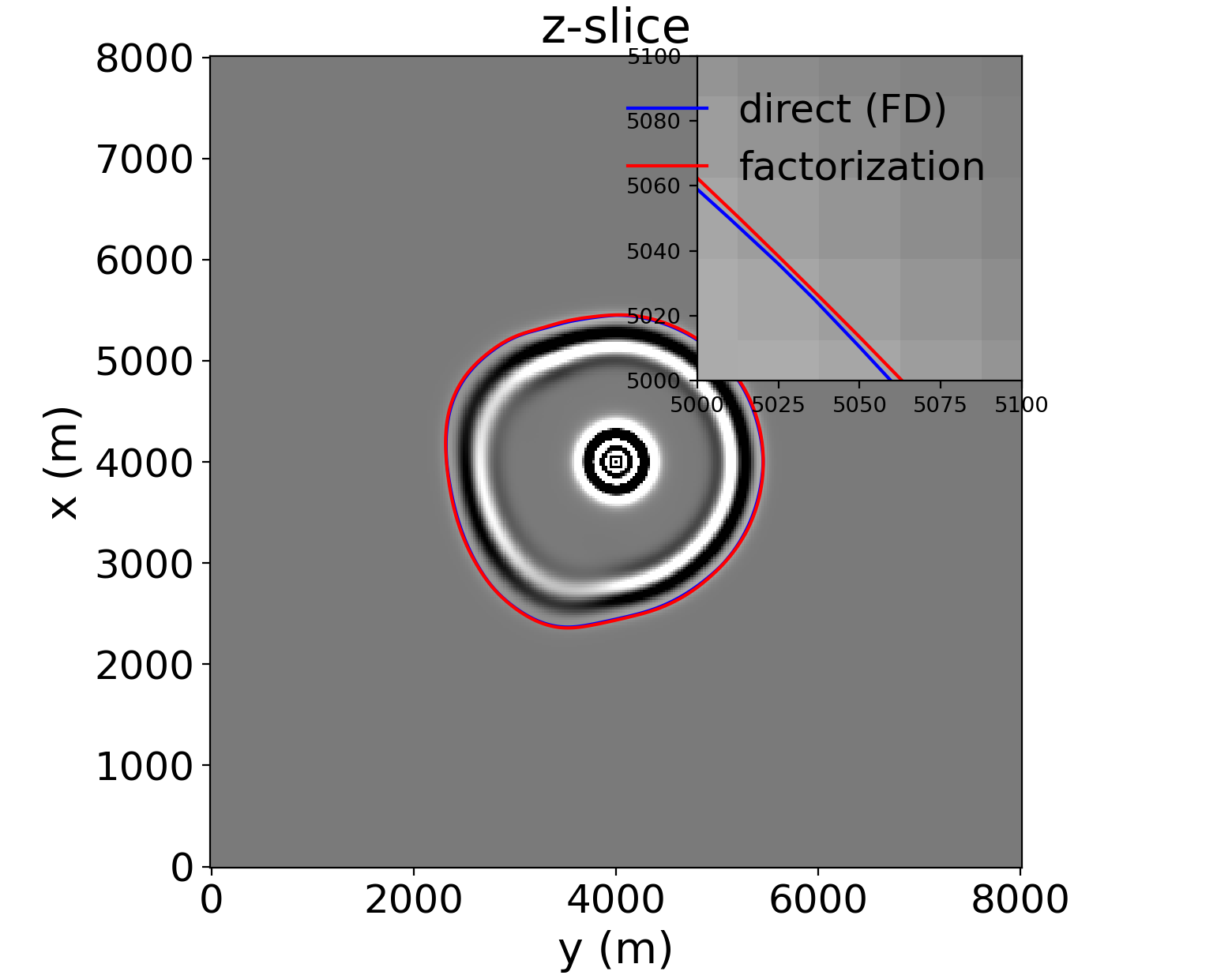}}
  \subfloat[]{\includegraphics[width=0.32\linewidth]{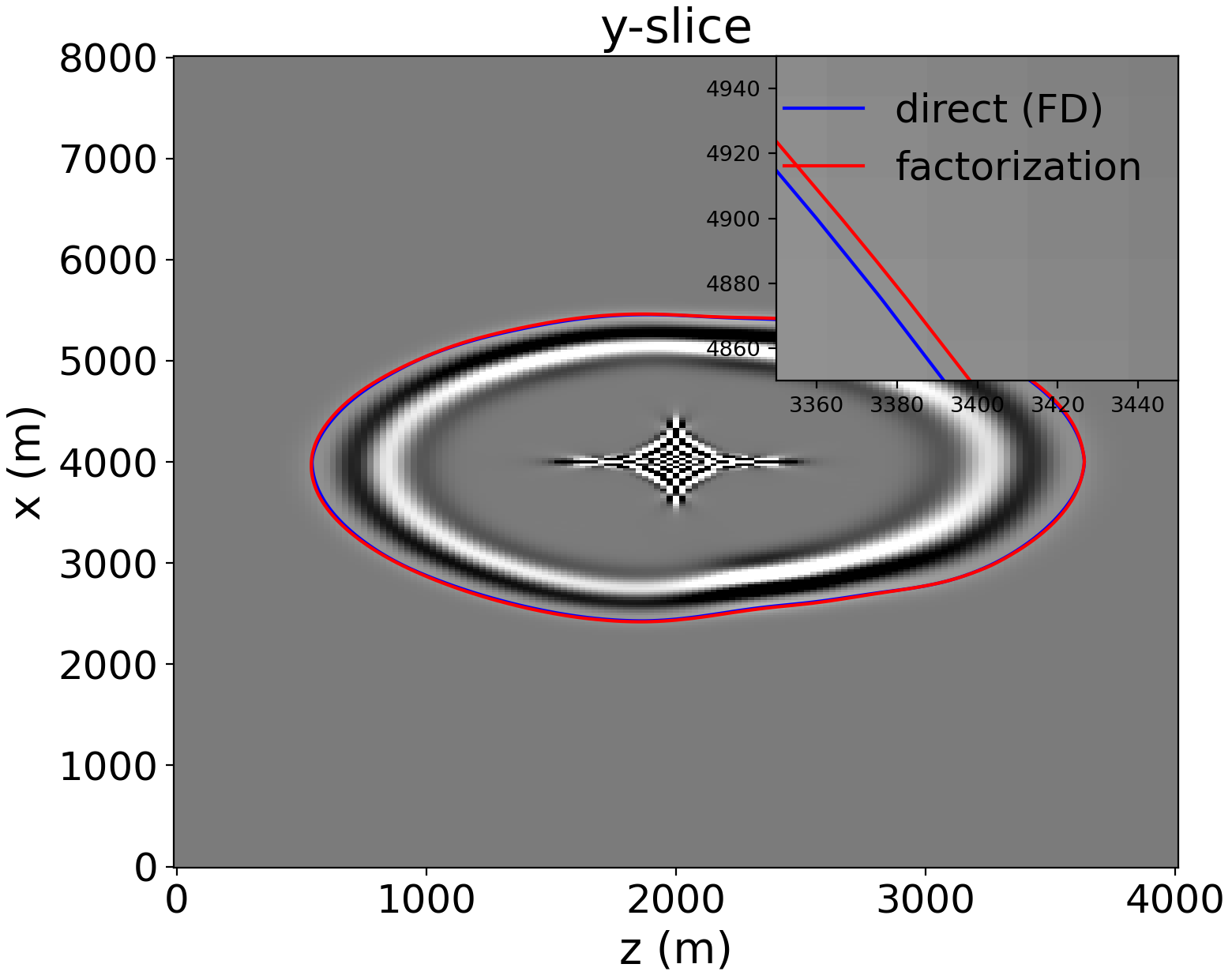}}
  \subfloat[]{\includegraphics[width=0.32\linewidth]{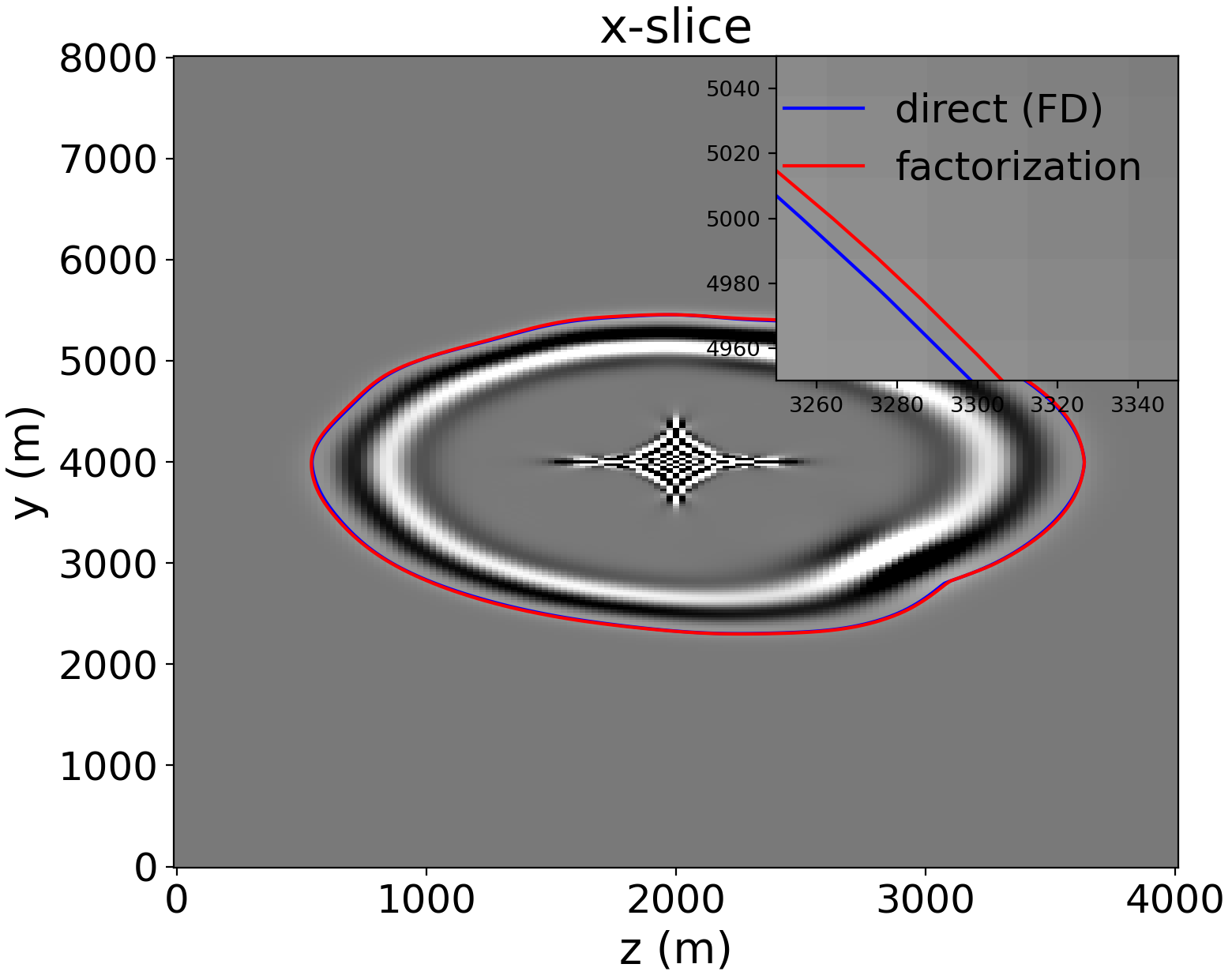}}
  \caption{Wavefield snapshot overlaid with traveltime contour $T=0.36$ s from direct FD (blue) and factorization (red).}\label{fig:over:wave_tt}
\end{figure}

We next examine our method based on the smoothed 3D Overthrust VTI model of size $321\times 321\times 161$ and grid spacings $dx=dy=dz=25$ m. The model parameters ($V_{p}$, $\epsilon$, and $\delta$) are shown in \Cref{fig:over:params}. 
The computed traveltime field on three orthogonal slices is presented in \Cref{fig:over:tt}. The wavefront geometry adapts naturally to the structural complexity of the model, without visible instability or nonphysical distortion. We validate the computed traveltime by overlaying the contour $T=0.36$ s on a pseudo-acoustic wavefield snapshot, as shown in \Cref{fig:over:wave_tt}. The close agreement between the contour and the simulated wavefront again demonstrates that the proposed solver captures first-arrival kinematics accurately in complex VTI media.

\begin{figure}[!htbp]
  \centering
  \subfloat[$\tau-1$]{\includegraphics[width=0.49\linewidth]{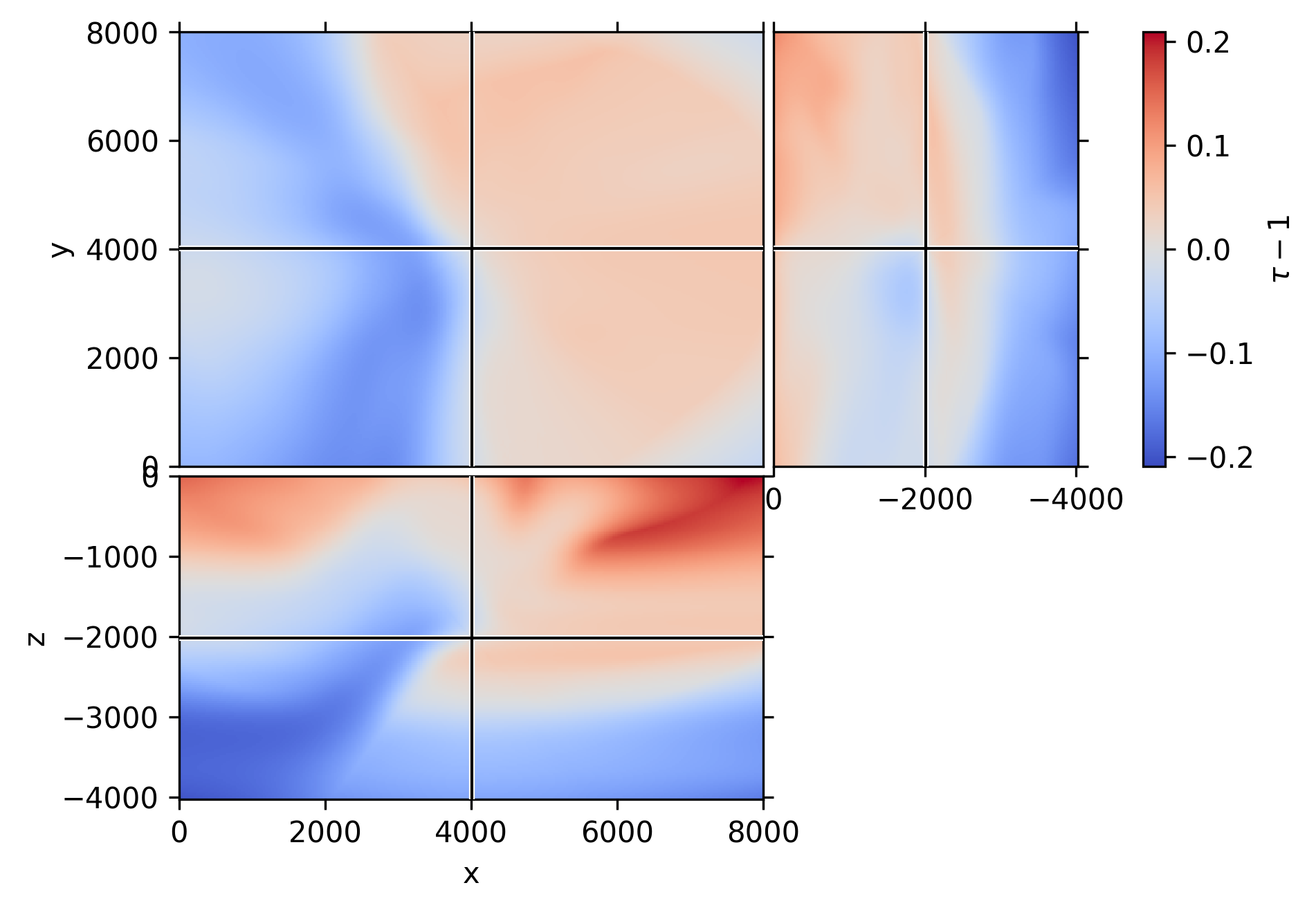}}
  \subfloat[$T-T_0$]{\includegraphics[width=0.49\linewidth]{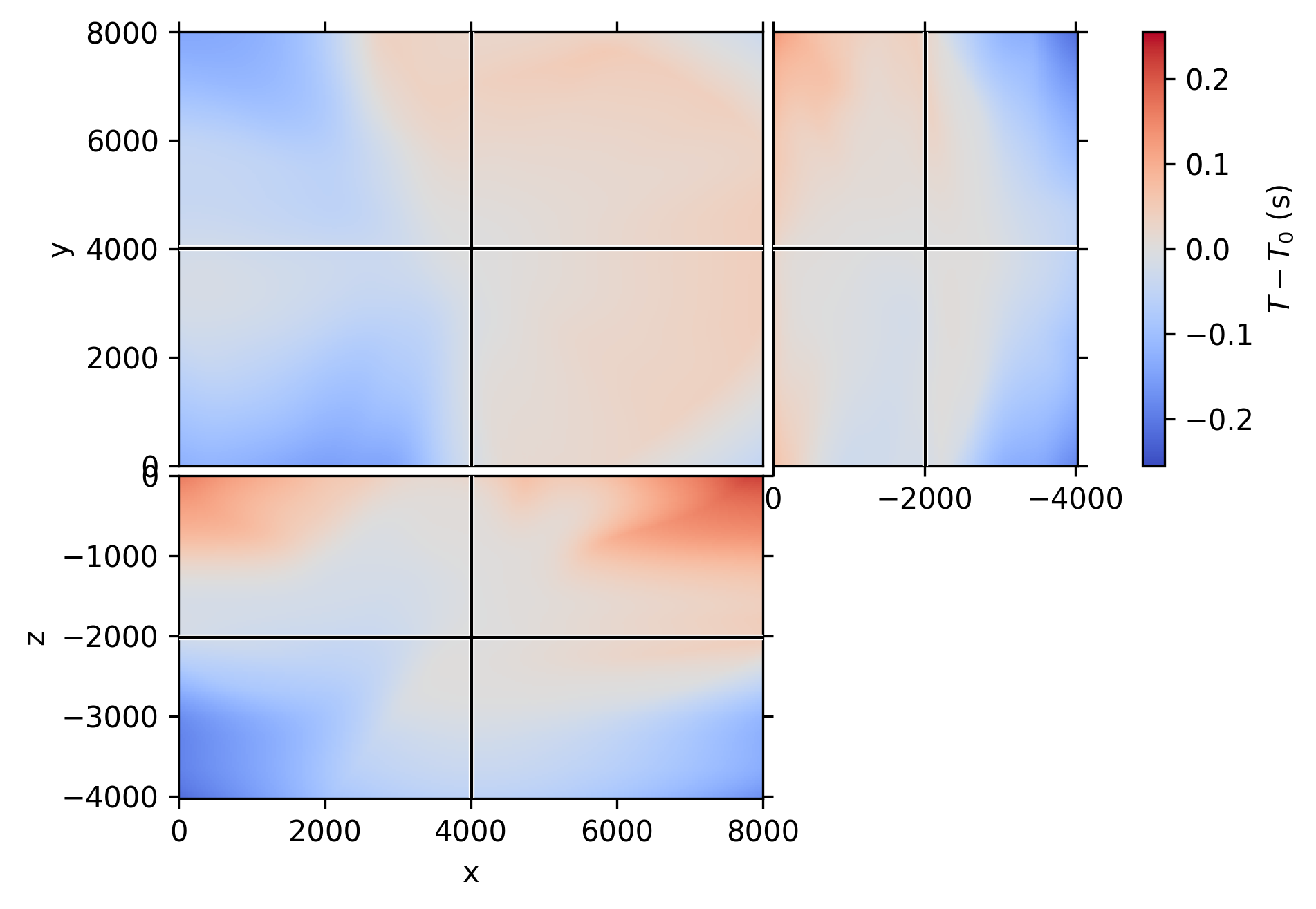}}
  \caption{Overthrust model: (a) The factorized correction field $\tau-1$; (b) The physical correction $T-T_0$ relative to the homogeneous reference field used in the factorization.}\label{fig:over:correction}
\end{figure}

The additional diagnostic in \Cref{fig:over:correction} highlights a different aspect of the method. In the Overthrust model, the correction field must absorb large-scale structural and anisotropic effects. The left panel of \Cref{fig:over:correction} shows that $\tau-1$ varies smoothly but nontrivially across the model, and the right panel shows the corresponding physical correction $T-T_0$, which again tracks the structural variations of the model. Quantitatively, the median, 90th-percentile, and 99th-percentile values of $|\tau-1|$ are $8.48\times 10^{-2}$, $1.88\times 10^{-1}$, and $2.23\times 10^{-1}$, respectively, with $\max |\tau-1| = 2.50\times 10^{-1}$ and $\max |T-T_0| = 3.42\times 10^{-1}$ s. These results show that the factorized formulation is not merely a near-source regularization device: in a strongly heterogeneous model it still produces a structured and stable correction field that carries the large-scale kinematic departures from the homogeneous reference.

\begin{figure}[!htbp]
  \centering
  \subfloat[$a$]{\includegraphics[width=0.32\linewidth]{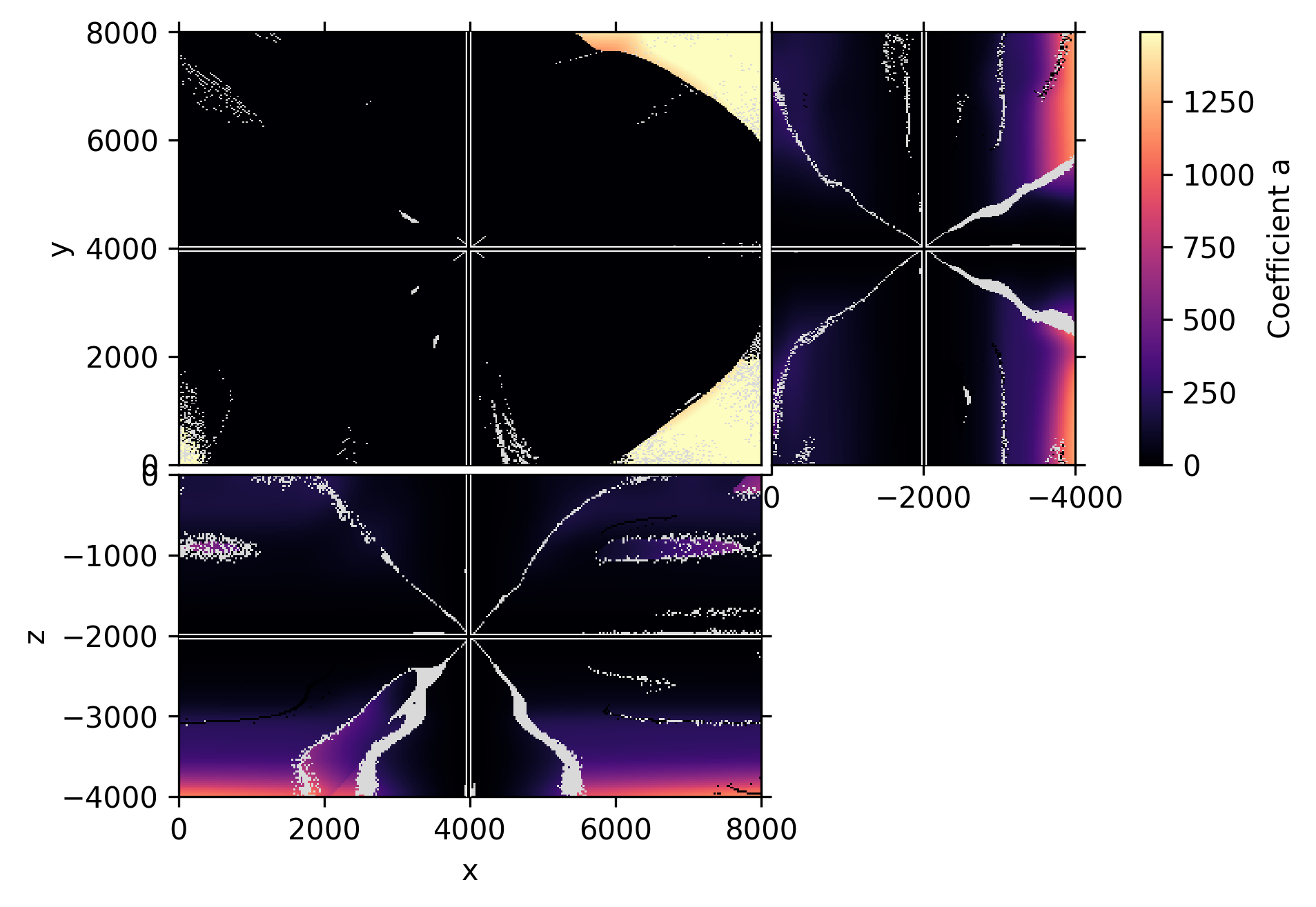}}
  \subfloat[$b$]{\includegraphics[width=0.32\linewidth]{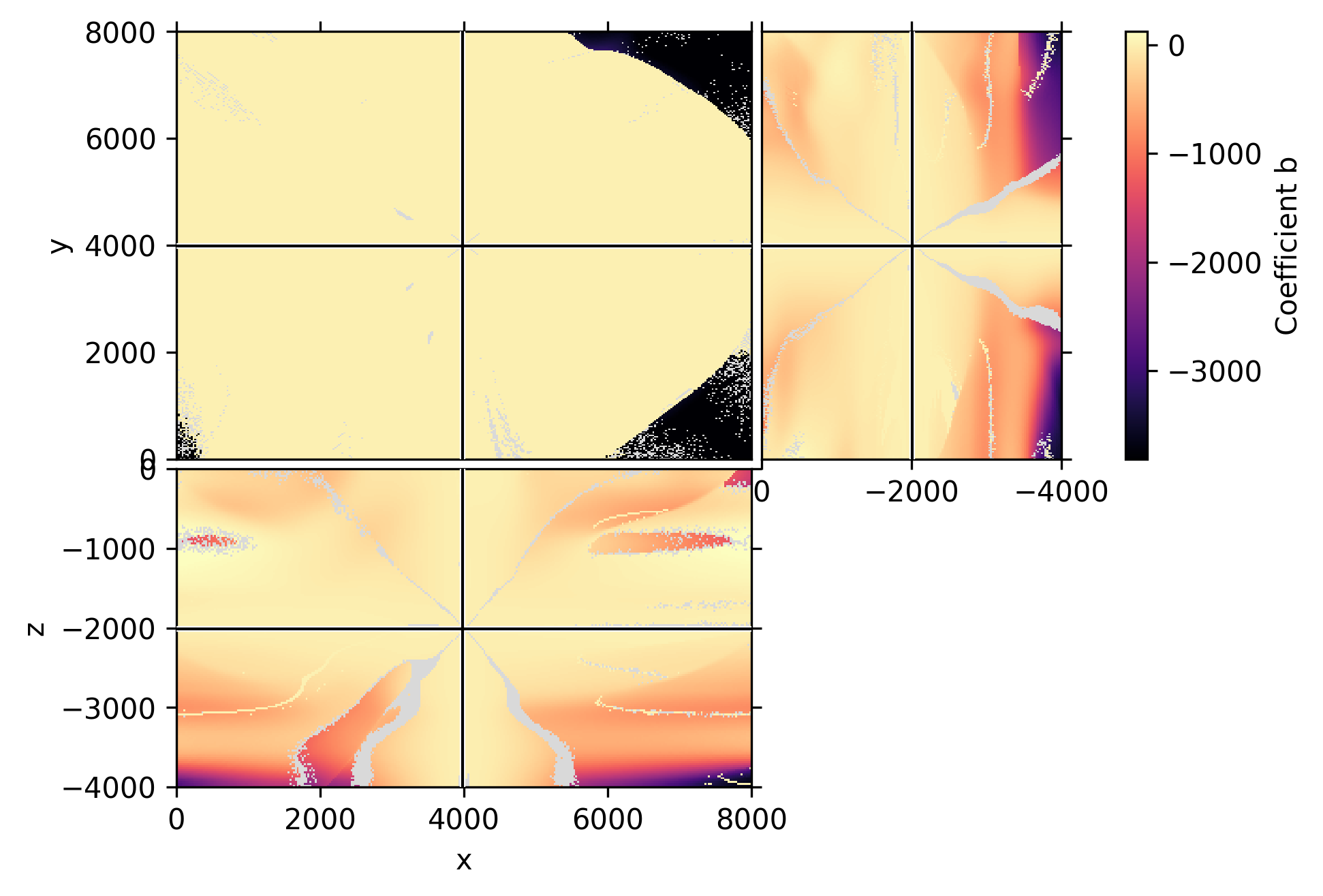}}
  \subfloat[$c$]{\includegraphics[width=0.32\linewidth]{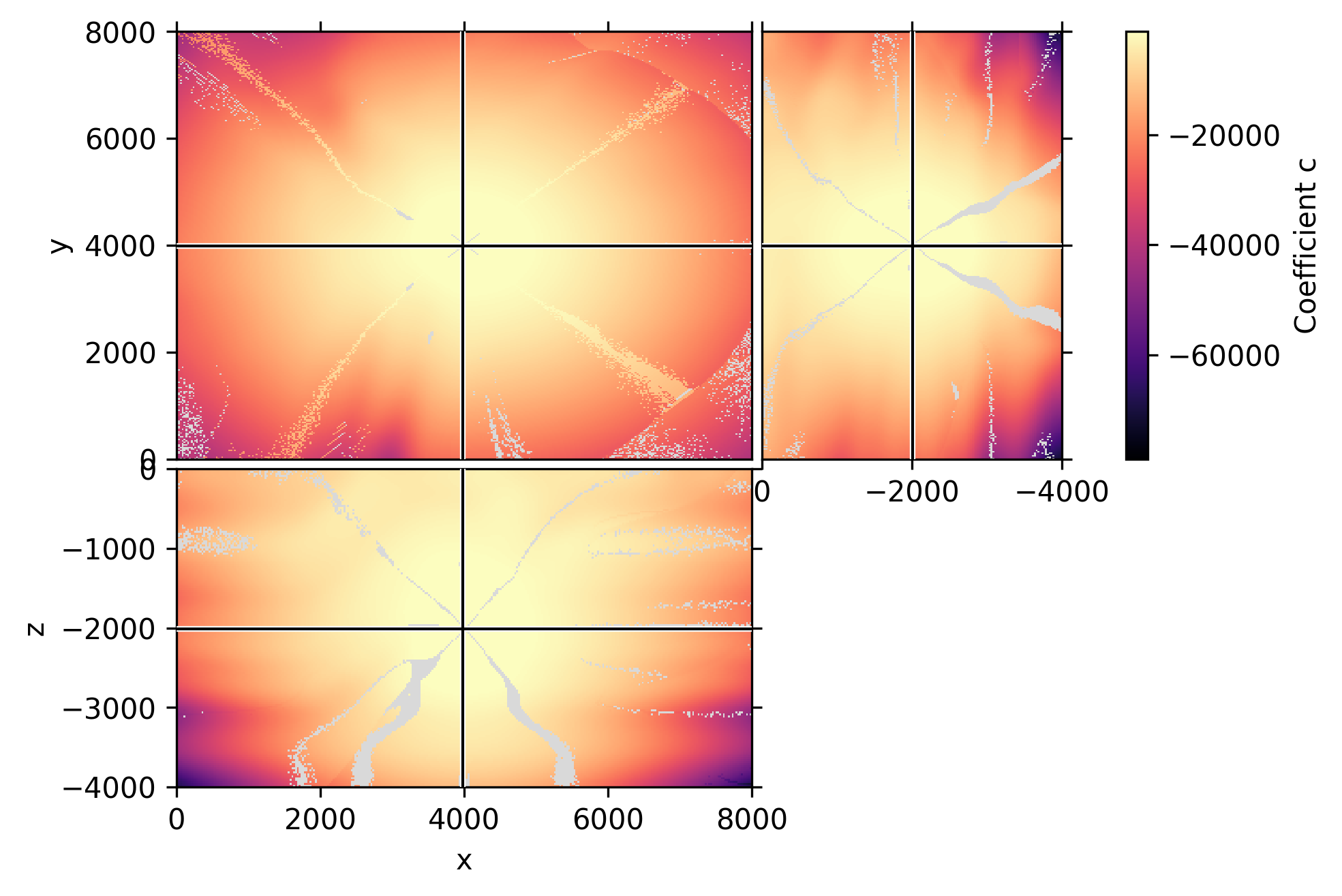}}\\
  \subfloat[$d$]{\includegraphics[width=0.32\linewidth]{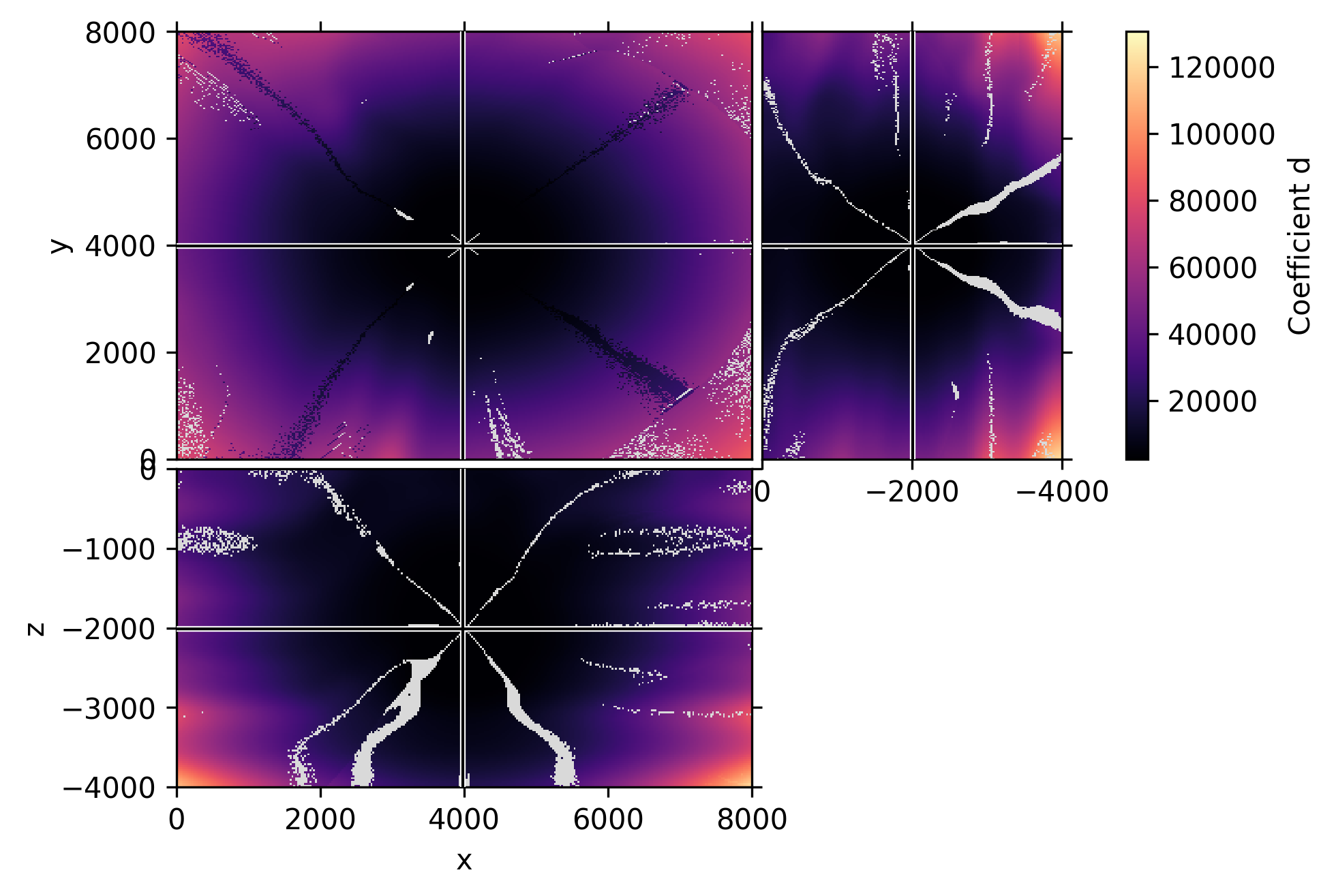}}
  \subfloat[$e$]{\includegraphics[width=0.32\linewidth]{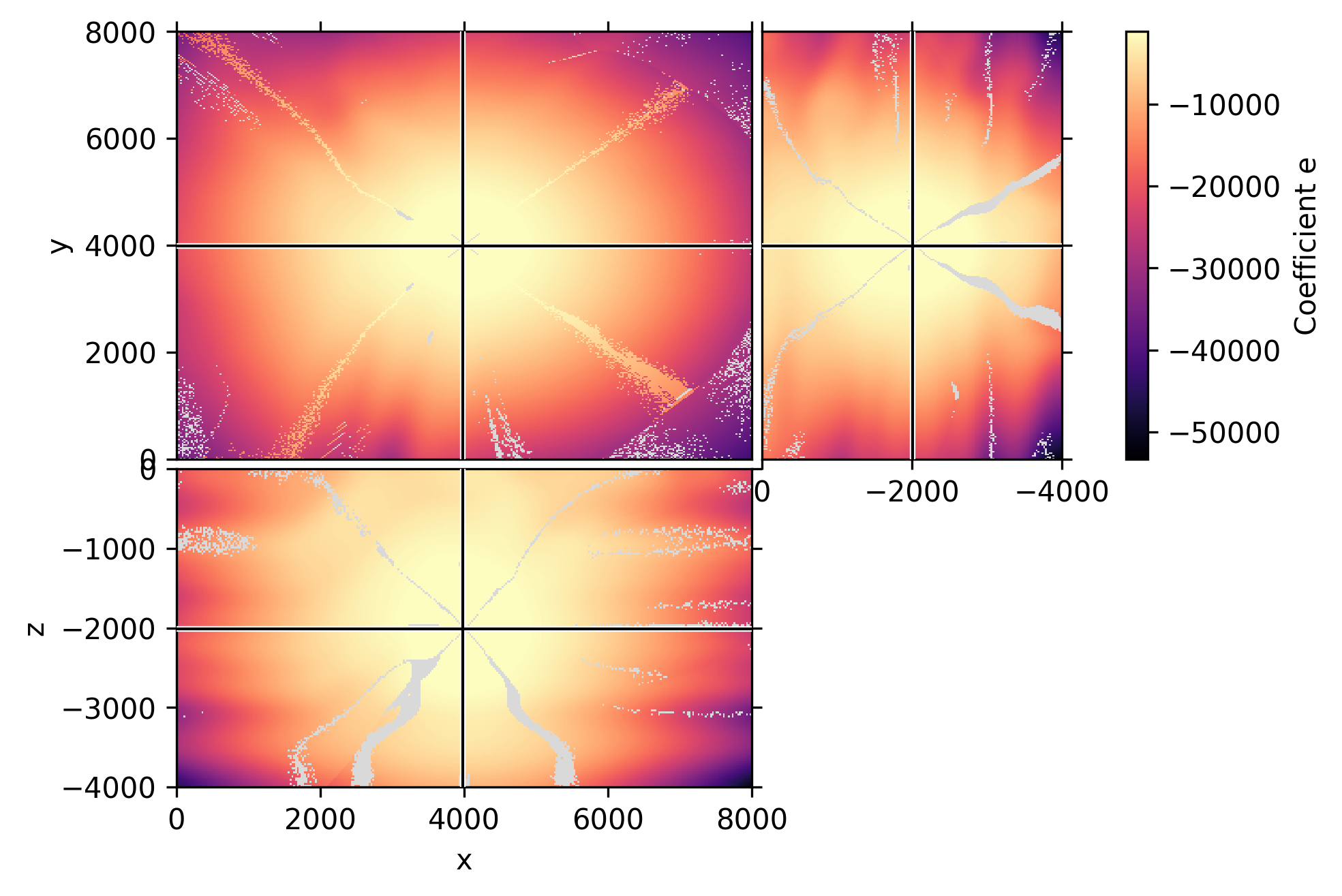}}
  \caption{The coefficients of the local quartic equation \eqref{eq:quartic} in the Overthrust model.}\label{fig:over:quartic}
\end{figure}

\Cref{fig:over:quartic} makes this branch dependence explicit for the Overthrust model. As in the homogeneous example, the coefficients $a$, $b$, $c$, $d$, and $e$ should be viewed as diagnostics of the accepted quartic branch rather than as globally defined solution variables. At each grid point, the solver accepts the smallest causal local update among several candidates, and only some of those candidates arise from the quartic branch. Consequently, the coefficient fields are defined only on the subset of cells where a quartic root is the accepted update; cells resolved by face-based or mixed-constraint fallback branches are rendered in gray.

Quantitatively, $16{,}530{,}434$ out of $16{,}589{,}601$ grid points select a quartic root as the accepted update in this Overthrust test. Over those quartic-update cells, the coefficient ranges are $a\in[0,\,4.16\times 10^{3}]$, $b\in[-8.55\times 10^{3},\,2.74\times 10^{2}]$, $c\in[-1.35\times 10^{5},\,-5.86]$, $d\in[7.84,\,2.14\times 10^{5}]$, and $e\in[-8.48\times 10^{5},\,-1.88]$. These values confirm that even in a strongly heterogeneous model the quartic coefficients remain bounded, while the support of the coefficient fields directly reveals where the full quartic branch is selected by the local update process.

\section{Conclusion}

We have developed a fast sweeping solver for the 3D VTI eikonal equation by combining multiplicative factorization with a six-tetrahedron pyramidal stencil. The proposed formulation removes the point-source singularity while retaining the improved local accuracy of the tetrahedral stencil. To handle the additional algebraic complexity introduced by factorization, we derive update strategies for both tetrahedral interiors and inclined faces, using Ferrari's method for quartic equations and a bisection-based procedure for the mixed constrained cases.
The numerical examples demonstrate the effectiveness of the proposed method. In the homogeneous VTI model, the factorized solver yields substantially smaller traveltime errors than the direct finite-difference method, while incurring only a moderate increase in computational cost. In the Overthrust model, the computed traveltime contours agree closely with pseudo-acoustic wavefield snapshots, indicating that the method captures first-arrival kinematics accurately in complex heterogeneous anisotropic media.

\section*{Acknowledgments}

Pengliang Yang thanks Prof. Jean Virieux for sharing his fast sweeping code and tutorial in Fortran to initiate this research. He was supported by National Natural Science Foundation of China (Grant No. 42274156). 
Yongming Lu was supported by the Key Project of the Spark Program of Earthquake Science and Technology in Hebei Province (DZ2025091100001) and University Stability Support Program of Shenzhen-General Project (20231128113233002). Jianming Zhang was supported by National Natural Science Foundation of China (Grant No. 42504112).   Deepseek was used to polish the grammar and spelling of this paper.

\section*{Data Availability Statement}

The source code and data are publicly available at: \verb|https://doi.org/10.5281/zenodo.19368293|.

\bibliographystyle{apalike}
\bibliography{ref.bib}

\end{document}